\documentclass[aps,prx,twocolumn,superscriptaddress,superscriptreference]{revtex4-2}

\usepackage{lmodern}
\usepackage[T1]{fontenc}
\usepackage[utf8]{inputenc}

\usepackage[english]{babel}
\usepackage[babel=true]{microtype} 
\usepackage{graphicx}
\usepackage{hyperref}
\usepackage{color} 

\usepackage{amsmath}
\usepackage{array}
\usepackage[table]{xcolor}
\usepackage{amsfonts}
\usepackage{amssymb}
\usepackage{bm}

\usepackage{placeins}

\usepackage{epstopdf}

\newcommand\equalhat{%
	\let\savearraystretch\arraystretch
	\renewcommand\arraystretch{0.3}
	\begin{array}{c}
		\stretchto{
			\scalerel*[\widthof{=}]{\wedge}
			{\rule{1ex}{3ex}}%
		}{0.5ex}\\ 
		=%
	\end{array}
	\let\arraystretch\savearraystretch
}

\newcommand{\diff}{\text{d}}
\newcommand\difft{\text{d} t}
\newcommand\kB{k_{\mathrm{B}}}

\newcommand\etaMD{\eta_\text{\tiny MD}}
\newcommand\etaMDL{\eta_\text{\tiny MDL}}
\newcommand\FMD{\tilde {F}_\text{\tiny MD}}
\newcommand\etaGLE{\eta_\text{\tiny GLE}}
\newcommand\FGLE{\tilde {F}_\text{\tiny GLE}}
\newcommand\gammaLPB{\tilde {\gamma}}

\newcommand{\gj}[1]{\textcolor{black}{#1}}
\newcommand{\bj}[1]{\textcolor{black}{#1}}

\begin{document}
	
	\selectlanguage{english}
	
	\title{Limits of the Non-Linear Generalized Langevin Equation: Cross-Correlations, Irreversibility and Desynchronization}
	
	\author{Bernd Jung}
\email{bernd@ralfj.de}
\affiliation{Independent Scholar, Eltville, Germany}

\author{Gerhard Jung}
\email{gerhard.jung.physics@gmail.com}
\affiliation{Institut f\"ur Theoretische Physik, Universit\"at Innsbruck, 6020 Innsbruck, Austria}

\date{\today}

\begin{abstract}

The generalized Langevin equation (GLE) is widely used to model complex soft-matter systems, including biomolecular dynamics, by incorporating memory effects and colored noise into coarse-grained descriptions. However, recent results suggest that combining memory with non-linear forces, ubiquitous in soft matter, introduces fundamental analytical inconsistencies. Here, using a simplified model, we investigate the practical numerical consequences of these analytical results \gj{in equilibrium systems}. We show that non-linear forces generate cross-correlations with the noise, modifying the fluctuation-dissipation theorem and rendering the noise position-dependent and irreversible. This implies that the commonly assumed reversible Gaussian noise in GLE simulations fails to capture essential features of the microscopic fluctuations. For weak non-linearities, these issues can be partially resolved either by using an iterative optimization of memory or by using microscopically consistent noise, which unexpectedly synchronizes GLE trajectories with the underlying microscopic dynamics. For stronger non-linearities like high barriers or shoulders in the external potential, however, iterative reconstruction fails and we observe desynchronization, indicating that the non-linear GLE no longer correctly reproduces the microscopic dynamics. Our results show in which situations non-linear GLEs can be accurately applied and when they fail, thus providing practical guidance for their application to coarse-grain soft-matter systems.

\end{abstract}
    
	\maketitle

	\section{Introduction}\label{sec:Intro}

	Almost 200 years ago, in 1828, Robert Brown discovered the random motion of small pollen particles in water \cite{Brown01091828}, now called Brownian motion. For quite a long time, the origin of the movement was not understood, until Albert Einstein in one of his 1905 manuscripts \cite{EinsteinBrown1905} showed that this phenomenon results from the sum of the kicks of a huge number of surrounding water molecules. He published the well known law for the mean-square displacement of the particles from their starting position, $\langle x^2(t) \rangle = 2Dt$, where $D$ is the diffusion coefficient. In his derivation, Einstein assumed that observable time scales are long so that the autocorrelation of velocity $v$ of the particle at different times is always zero. This is called the diffusive limit.
	
	Only three years later, Paul Langevin developed a stochastic differential equation, based on heuristic considerations \cite{Langevin1908},
	\begin{equation}\label{eq:LangEq}
		F(t) = m \frac{\diff v}{\difft} = F_\text{e}(t) - \gamma v(t) + \eta(t).
	\end{equation}
	Here $F$ is the total force acting on the particle, $m$ is its mass, $F_\text{e}$ an external force, $\gamma$ a friction coefficient, and $\eta$ is a fluctuating force modeled as Gaussian white noise obeying the fluctuation-dissipation relation (FDR) $\langle \eta(t)\eta(t') \rangle = 2 \gamma \kB T \delta(t-t')$. In this Langevin equation (LE), the forces $\tilde{F}(t)$ of the environment are replaced by the sum of the systematic dissipative term $- \gamma v(t)$ and the noise. For an overdamped particle in thermodynamic equilibrium, the LE reproduces Brownian motion: $\langle x^2(t) \rangle = 2 \frac{d \kB T}{\gamma} t$, where $d$ is the dimension of space. Without this condition, however, Eq.~(\ref{eq:LangEq}) considers inertia of the particle, displaying a non-zero velocity autocorrelation function (VACF) decaying exponentially with time,
	\begin{equation}\label{eq:v2}
		\langle v(t)v(t') \rangle = \frac{\kB T}{m} \text{e}^{-\frac{\gamma}{m} |t-t'|}.
	\end{equation}
	
	The LE is very useful for computer simulations, because it allows modeling the motion of molecules without explicitly simulating the tremendous amount of surrounding ``bath'' particles, thus resulting in a formidable acceleration of computation. Such a procedure is called dynamic coarse-graining and is used in many areas of soft matter, in particular for polymers \cite{hess2006long,harmandaris2007ethylbenzene,padding2011systematic,zhang2025topological}, proteins \cite{best2011diffusion} and membranes \cite{deserno2009mesoscopic}. On the time scale at which correlations in the bath are decaying, however, the LE does not describe the VACF correctly since the coarse-grained dynamics is simply Markovian: it ignores any correlations within the environment. In particular in complex soft-matter systems with viscoelastic environment, the Markovian assumption therefore needs to be critically regarded \cite{KlippJung2021,dalton2023fast,koch2024analysis,motahari2025chapman}.
	
	This shortcoming was cured when Robert Zwanzig and Hajime Mori introduced the projection operator formalisms to derive the linear generalized Langevin equation (GLE) \cite{Zwanzig1961, Mori1965, Zwanzig2001,vrugt2020projection}
	\begin{equation}\label{eq:GLE}
		F(t) = m \frac{\diff v}{\difft} = F_\text{e}(x(t)) - \int_{0}^{t} K(t-s)v(s) \diff s + \eta(t)
	\end{equation}
	together with the corresponding FDR \cite{kubo1966fluctuation, hanke2025second},
	\begin{equation}\label{eq:FDR}
		\langle \eta(0)\eta(t) \rangle = \kB T K(t).
	\end{equation}
	\gj{Here,} $F_\text{e}$ is a linear external force, and $K(t)$ is the memory kernel. The GLE \bj{exhibits} two main differences to the LE. First, the dissipative force is now replaced by a convolution of the time-dependent kernel $K(t)$ with the history of $v(t)$ instead of a simple instantaneous friction term. Second, $\eta(t)$ is not $\delta$-correlated white noise anymore, but instead the fluctuating force $\eta(t)$ exhibits non-trivial correlations and therefore needs to be modeled as colored noise. Nevertheless, this noise is still assumed to be Gaussian. The non-Markvian GLE (\ref{eq:GLE}) is able to reproduce VACFs because it models complex interactions between the coarse-grained variables and the environment.
	
	In the original version of the GLE as derived by Mori and Zwanzig (MZ GLE), the external force is inherently linear. Nevertheless, GLEs are often used in combination with non-linear external or conservative forces, in particular to model soft matter and biomolecular dynamics \cite{lange2006collective,kinjo2007equation,li2017computing,JungHankeSchmid2018,Zaccone_viscosity_2023,zhang2024chemically,ge2024data}. This has sparked an intensive analytical debate on the usage of non-linear forces in projection operator techniques \cite{glatzel2021interplay,vroylandt2022position,Vroylandt2022,JungJung2023,izvekov2025mori,plyukhin2026langevinpmf,plyukhin2026langevin,Koch_2026}. The gist of this discussion is that a simple GLE as presented in Eqs.~(\ref{eq:GLE}) and (\ref{eq:FDR}) with non-linear external forces cannot be rigorously derived and features position-dependent memory kernels \cite{vroylandt2022position} or other additional non-trivial terms \cite{glatzel2021interplay}.    In particular, Vroylandt \cite{Vroylandt2022} has shown \gj{using projection operators} that combining non-linear external forces with a position-independent memory kernel leads to a violation of the FDR in Eq.~(\ref{eq:FDR}). Instead, the FDR contains additional terms corresponding to cross-correlations between the external force and the colored noise (\gj{see also Ref.~\cite{shea2024force} for a derivation without projection operators}). Neglecting these cross-correlations leads to incorrect dynamics \cite{Klipp2021}.  Numerically, it has been shown that this deficiency can be overcome by modifying the memory kernel via an approximation formula or via iterative reconstruction such that the VACF is reproduced correctly \cite{Klipp2021, klippenstein2023bottom, Klippensteinetal2024, Wolf2025, wolf2025matching}.

    Based on a simple model system reminiscent of the Caldeira-Leggett model we have recently presented analytical and numerical results on the impact of introducing non-linearities to GLE models \cite{JungJung2023}. In particular, we have shown that higher-order correlation functions are not correctly reproduced by the GLE. In all systems in Ref.~\cite{JungJung2023}, however, either the external force or the tagged particle-oscillator coupling was linear.  In this work, we extend this approach to fully non-linear models which feature dynamics very reminiscent of biomolecular systems, including complex correlations within the environment leading to significant memory effects. However, the simplicity of the model allows us to control the non-linearity in the system and thus systematically investigate how it leads to the emergence of non-trivial correlations between the external and the other forces.
    
    After introducing the system and modeling techniques in Sec. \ref{sec:Methods}, we first investigate the precise properties of the fluctuating force $\eta(t)$, which is an essential input for the GLE in Sec. \ref{sec:Analysis}. We show that already in the absence of external or conservative non-linear forces $\eta(t)$ is irreversible. When introducing non-linearities, $\eta(t)$ features further non-trivial properties, such as cross-correlations with the external force which enter the FDR, precisely as predicted in Ref.~\cite{Vroylandt2022}, and a non-linear position-dependence of the fluctuations. All of these properties are, of course, lost when modeling $\eta(t)$ as a simple colored noise as is usually done in GLE simulations. Second, in agreement with Refs.~\cite{klippenstein2023bottom} we find in Sec.~\ref{sec:Iterative} that an iterative optimization can be used to find non-linear GLEs which reproduce the correct VACF. However, for external potentials with a high potential barrier ($> \kB T$) we find that the iterative reconstruction does not converge anymore and features memory kernels with negative spectrum.  
    
    Third, we study the dynamics of non-linear GLEs which receive as input the fluctuating force directly extracted from microscopic simulations. We introduce this technique in Sec. \ref{sec:GLELPB} and call it \textit{playback} since it is based on existing trajectories and basically tends to recreate the original dynamics as precisely as possible. The method has also recently been referred to as 'non-Gaussian orthogonal forces' \cite{Netz2025}. Here, we show that it not only correctly reproduces the microscopic correlation functions, but actually leads to synchronization in the sense that - independent of initial conditions - the GLE will ultimately replicate exactly the same microscopic trajectories. When introducing external potentials with high barriers ($> \kB T$), however, we observe desynchronization, i.e., even if the GLE has the same initial conditions as the microscopic simulation, trajectories and relevant average quantities will eventually become completely different and even the Boltzmann distribution is not correctly reproduced. Fourth, we investigate the response of the system to small external perturbations to study how well the GLE model can reproduce the transport properties of the system.

	\section{Model and Methods}
	\label{sec:Methods}
	
	In this paper, we consider the same stochastic non-linear Caldeira-Leggett (SNCL) model which we analyzed in our previous manuscript \cite{JungJung2023}. It \gj{describes the equilibrium dynamics} of a single tagged particle in a one-dimensional external potential, which is usually quartic,
	\begin{equation}\label{eq:extPot}
		V_\text{e}(x_0) = \frac{1}{2} a_\text{e} x_0^2 + \frac{1}{4} b_\text{e} x_0^4.
	\end{equation}
    We mainly study three different external potentials: (i) harmonic ($b_{\text{e}} = 0$), (ii) anharmonic ($a_{\text{e}}>0, b_{\text{e}} >0$), or (iii) a potential barrier around $x_0=0$ with barrier height $h = \frac{1}{4} a_\text{e}^2/b_\text{e}$ ($a_{\text{e}}< 0$) . If other types of external potentials are applied, this will be specified.
	The tagged particle is coupled to one or more ($i=1, ..., N$) oscillators via the potential,
	\begin{equation}\label{eq:OscPot}
		V_i(\Delta x_i) = \frac{1}{2} a_i (\Delta x_i)^2 + \frac{1}{4} b_i (\Delta x_i)^4,
	\end{equation}
	where $\Delta x_i = x_0 -x_i$. $x_0$ and $x_i$ are the positions of particle and oscillators, respectively. The oscillators are subject to their coupling potential $V_i$ only but not to $V_{\text{e}}$. They are connected via Langevin dynamics with friction coefficient $\gamma$ to a heat bath as described in Appendix A of Ref.~\cite{JungJung2023}. These oscillators are modeling complex interactions between the tagged particle and the bath, for example in complex fluids \cite{Kupferman2004,wang2019implicit}.
    
    \subsection{Molecular dynamics (MD) simulations}
    
    The dynamics of these systems is integrated via molecular dynamics (MD) simulations as described in \cite{JungJung2023},
	\begin{align}\label{eq:eomMD1}
		\frac{\diff}{\difft} x_0 &= v_0\\
		m_0\frac{\diff}{\difft} v_0 &= F_0(t) = - V_{\text{e}}^\prime(x_0(t)) + \FMD(t) \\
		\FMD(t) &= - \sum_{i=1}^N V_i^\prime (\Delta x_i(t)) \\
		\frac{\diff}{\difft} x_i &= v_i \\
		m_i\frac{\diff}{\difft} v_i &= V_i^\prime (\Delta x_i(t)) - \gamma v_i(t) + W_i(t), 
	\end{align}
	where $W_i(t)$ is white noise with $\langle W_i(t) W_i(t^\prime) \rangle = 2 \gamma \kB T \delta(t-t^\prime)$. The oscillators therefore play the role of a complex environment to which the tagged particle is coupled. For example, our model can describe the motion of macromolecules or colloids in complex fluids \cite{JungJung2023}. We use $\kB T$ in units of energy $\epsilon$, mass in units of $m$, time in units of $\tau$, $\gamma$ in $m \tau^{-1}$, and length in $\sigma = \tau \sqrt{\epsilon / m}$.
	
	Simulation time steps were chosen as $\Delta t = 0.005 \tau$ and the number of oscillators $N=1$ throughout this work, which is sufficient to observe complex behavior. The mass of the particle is $m_0 = 1m$ unless otherwise specified. \gj{We performed MD simulations for at least $N_t = 8.4 \cdot 10^7$ time steps. We validated that sampling was sufficient to ensure that statistical error bars would be much smaller than the data point size in the following figures.}. Investigated systems are listed in Table \ref{tab:systems}.
	\begin{table}[h!]
		\centering
	\begin{tabular}{c|c c c c c c c c}
    		\hline
		 System & $\dfrac{ a_\text{e}}{\epsilon \sigma^{-2}}$ & $\dfrac{b_\text{e}}{\epsilon \sigma^{-4}}$ & $\dfrac{a_1}{\epsilon \sigma^{-2}}$ & $\dfrac{b_1}{\epsilon \sigma^{-4}}$ & $\dfrac{m_1}{m}$ & $\dfrac{\gamma}{m \tau^{-1}}$ & $\dfrac{\kB T}{\epsilon}$  \vspace{0.05cm}\\

		\hline

		FH & 0 & 0 & 4.6781 & 0 & 1 & 5 & 1 \\

        FA & 0 & 0 & 0 & 10 & 1 & 5 & 1 \\

        HA & 5.1234 & 0 & 0 & 13 & 2 & 15 & 2 \\
		
		BH & -10 & 6 & 10.34 & 0 & 2 & 15 & 2 \\
		
		B$n$A & -$n$ & 6 & 0 & 13 & 2 & 15 & 2 \\

        		\hline
		
	\end{tabular}
    		\caption{Investigated systems and their parametrization ($m_0/m = 1$). From top to bottom: (i) free tagged particle, harmonically coupled (FH), (ii) free tagged particle,  anharmonically coupled (FA), (iii) harmonically confined tagged particle, anharmonically coupled (HA), (iv) external potential barrier, harmonic coupling (BH), (v) external potential barrier of height $h/\epsilon = n^2 / 24$, anharmonic coupling (B$n$A).}
		\label{tab:systems}
	\end{table}

    \subsection{Generalized Langevin dynamics (GLE) simulations}

    The goal of the coarse-graining procedure is to integrate out the complex environment and replace it by an effective description where the dynamics is modeled via the GLE with memory and colored noise. This procedure is standard and has been reviewed, e.g., in Ref.~\cite{KlippJung2021}. In this manuscript, GLE simulations are conducted in the same way as described in Ref.~\cite{JungJung2023} by numerically integrating the equations,
	\begin{align}\label{eq:eomMD2}
		\frac{\diff}{\difft} x_0 &= v_0\\
		m_0\frac{\diff}{\difft} v_0 &= F_0(t) = - V_{\text{e}}^\prime(x_0(t)) + \FGLE(t) \\
		\FGLE(t) &= {F}_\text{d}(t) + \etaGLE(t) \\
        {F}_\text{d}(t) &= - \int_{-\infty}^{t} K(t-s)v_0(s) \diff s .
	\end{align}
    \gj{Here, $- V_\text{e}^\prime(x_0(t)) = F_\text{e}(x_0(t))$ is the mean force acting on the coarse-grained particle, which corresponds in our case to the original external force.}
    Effectively, the GLE therefore attempts to approximate the particle-bath coupling $\FMD$ by $\FGLE$, which is the \bj{sum} of a dissipative force ${F}_\text{d}(t)$ and a fluctuating force $\etaGLE.$ Importantly, as is commonly done for GLE simulations, we consider the stationary version of the GLE (\ref{eq:GLE}) by integrating starting from the infinite past instead of having a non-stationary dynamics starting at $t=0.$ In practice, we introduce a cutoff as the maximal time at which memory effects are considered.
	 If not otherwise specified, we use the Volterra-Kernel $K_{\text{V}}$ calculated from correlation functions extracted via the above described MD simulations \gj{by inverting the Volterra equation of second kind \cite{shin2010brownian}}.  The inverse algorithm in the presence of external forces is \cite{JungJung2023},
	\begin{align}\label{eq:Kernel}
		K_{\text{V}}(i  \Delta t) &= \frac{1}{\langle v_0^2 \rangle} \left\{ \frac{1}{m_0} \langle F_0(0) \FMD(i \Delta t) \rangle - \right. \\ &\left. \frac{\Delta t}{m_0} \sum_{j=0}^{i-1} w_j \langle v_0(0) {F_0}( (i-j)\Delta t) \rangle K_{\text{V}}(j  \Delta t)  \right\}, \nonumber
	\end{align}
	with initial condition $K_{\text{V}}(0) = \frac{\langle F_0(0) \FMD(0) \rangle}{m_0 \langle v_0^2 \rangle }$ and weight factor $w_j$ with $w_0 = 1/2$ and $w_j = 1$ for $j > 0$.    
    
    As commonly done in GLE simulations, the fluctuating force $\etaGLE$ is modeled as colored, i.e., correlated Gaussian noise with $\langle \etaGLE(0)\etaGLE(t) \rangle = \kB T K(t).$ These random numbers are created by the Fourier transform technique described in detail in Appendix A of Ref. \cite{JungHankeSchmid2018}.  \gj{ In the GLE simulations we do not consider complex properties of the fluctuating force such as non-Gaussian distributions \cite{JungJung2023}, its position dependence or its irreversibility. The implications of this approximation will be analyzed in the next section.}

	\section{Analysis of {colored} noise and {non-Markovian} dissipative force}
	\label{sec:Analysis}

    When modeling complex dynamics using the GLE, the total force is separated into three different contributions as described after Eq.~(\ref{eq:GLE}). In particular, the fluctuations induced by seemingly random collisions with the bath particles are replaced by noise modeled as correlated random numbers. This step is crucial since it allows to generate coarse-grained trajectories on arbitrary time scales without knowledge on the precise motion of the bath particles. However, as we will show in this section, replacing the microscopic fluctuations by noise is a very subtle procedure and introduces critical approximations in particular in non-linear GLEs.
	
	\subsection{What is noise?}
	\label{sec:WhatIsNoise}

    Already the answer to the question 'what is noise' depends on the point of view. When deriving the GLE using the MZ projection operator formalism, the fluctuating force $\eta(t)$ is defined as \bj{the sum of} all forces which are orthogonal to the coarse-grained variables and which thus depend solely on the initial conditions of the bath particles \cite{Zwanzig2001}. Starting from deterministic microscopic dynamics, the fluctuating force is therefore also deterministic. Only a posteriori it can be interpreted as noise, assuming that we do not know the initial conditions and thus choose a random realization from the microscopic distribution function.
    
    Here, we will take a complementary perspective. Starting from the GLE as a stochastic model to describe complex dynamics, we reconstruct the memory from MD trajectories as described in the previous section. Subsequently, this allows us to define the microscopically consistent fluctuating force $\etaMD(t)$ as all forces remaining after subtracting conservative \bj{($F_\text{e}$)} and dissipative forces from the total force \bj{($F_0$)} measured in MD simulations,
    \begin{align}\label{eq:MDnoise}
		\etaMD(t) &= F_0(t) - F_\text{e}(x(t)) + \int_{-\infty}^{t} K(t-s)v_0(s) \diff s.
	\end{align}
    This approach allows us to extract the fluctuating force $\etaMD(t)$ and analyze its properties, even in systems with non-linear $F_\text{e}(x(t))$. 
    
    \gj{As already shown in Ref.~\cite{JungJung2023} if the tagged particle-oscillator coupling $\tilde F_\text{MD}$ is non-linear we find that the fluctuating force $\etaMD$ becomes non-Gaussian. In the following, we will investigate further complex properties of $\etaMD$. }

    \subsection{Irreversibility of noise}
	\label{sec:IrreversNoise}

    	\begin{figure}
        \includegraphics[width=0.95\linewidth]{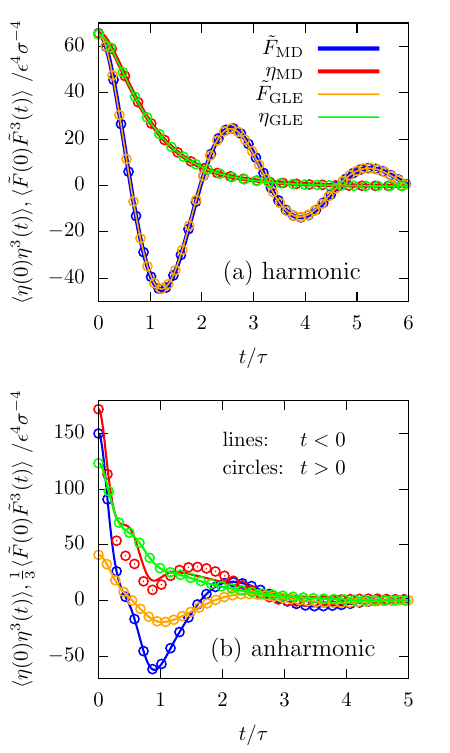}
		\caption{Time-reversibility of the fluctuating force $\etaMD$ analyzed via the correlation functions $\langle \eta(0)\eta^3(t) \rangle$ and $\langle \tilde{F}(0)\tilde{F}^3(t) \rangle$ from MD and GLE simulations for two systems without external potential. (a) Purely harmonic system (FH), (b) anharmonic oscillator coupling (FA).}
		\label{fig:FigSKO}
	\end{figure}

     It is known that coarse-graining can render dynamics irreversible \cite{robertson2020asymmetry}. Therefore, as a first analysis we applied a method as described in Ref. \cite{lucente2023statistical} to check the hidden irreversibility of a quantity $\text{X}(t)$: we examined the 'odd' fourth-order correlation $\langle \text{X}(0) \text{X}^3(t) \rangle$ for $\text{X} = \eta$ and $\text{X} = \tilde{F}$. Fig.~\ref{fig:FigSKO} shows the results of MD and GLE simulations \textit{without} external potential. It can be seen that $\tilde{F}(t)$  is always time-reversible, which is, of course, a direct consequence of the time-reversibility of the microscopic dynamics.
    However, when the coupling between the tagged particle and the oscillator  is non-linear, as is expected in general soft-matter systems, $\etaMD(t)$ becomes irreversible (see Fig.~\ref{fig:FigSKO}(b)). This is an important difference between the noise in MD simulations and the one usually used in GLE simulations, which is perfectly reversible. Importantly, no external or potential or conservative interactions are needed to observe this effect. It is really an inherent property of the non-Markovian noise $\etaMD$, after splitting of the dissipative force $F_\text{d}(t)$. \gj{In particular, the irreversibility is not linked to a violation of the FDR (see Appendix \ref{app:ValidFDR}) or other properties analyzed below for non-linear GLEs.}

\begin{figure}
	\includegraphics[width=0.95\linewidth]{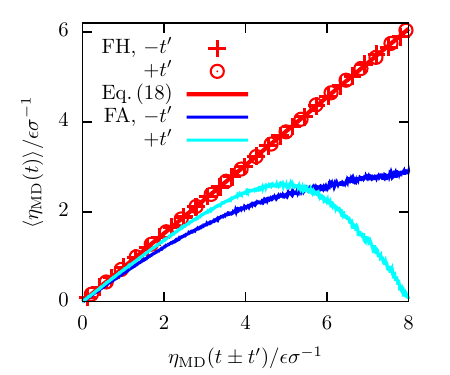}
	\caption{Explicit visualization of the time-irreversibility using the conditional average of $\etaMD(t)$ as a function of $\etaMD(t \pm t^\prime)$ ($t^\prime = 0.455 \tau$) for the same systems as in Fig.~\ref{fig:FigSKO}. The straight red line is calculated according to Eq.~(\ref{eq:condmean}).}
	\label{fig:Figetalag}
\end{figure}

This irreversibility can also be studied more explicitly, as elucidated in Fig.~\ref{fig:Figetalag} and Appendix \ref{app:DerCond}. If the oscillator is coupled harmonically to the particle, the conditional mean value of the noise at time $t$, $\langle \etaMD(t) \rangle$, depends linearly on the noise history at time $t - t^\prime$, and this dependence is identical if we plot $\langle \etaMD(t) \rangle$ as a function of the future noise at $t + t^\prime$. Furthermore, the variance of $\etaMD$ at time $t$ is constant and does not depend on $\etaMD$ at $t - t^\prime$ or $t + t^\prime$. We can derive (cf. Appendix \ref{app:DerCond}) simple equations for these relations, which make it possible to calculate the conditional values just by knowledge of the kernel,
\begin{align}\label{eq:condmean}
	\langle \etaMD(t) \rangle = \mathfrak{a} \, \etaMD(t \pm t^\prime) \, \text{ with } \, \mathfrak{a} = K_{\text{V}}(t^\prime) / K_{\text{V}}(0)
\end{align}
\begin{align}\label{eq:condvar}
	\sigma_{\etaMD(t)}^2 = \mathfrak{c}^2 \, \text{ with } \, \mathfrak{c}^2 = \kB T K_{\text{V}}(0) (1 - \mathfrak{a}^2).
\end{align}
Eq.~(\ref{eq:condmean}) fits perfectly to the numerical results in Fig.~\ref{fig:Figetalag}, where we have chosen $t^\prime = 0.455 \tau$, because for this value the irreversibility is most pronounced in Fig.~\ref{fig:FigSKO}(b).

If, however, the oscillator is coupled anharmonically to the particle, then the dependence of $\langle \etaMD(t) \rangle$ on $\etaMD(t \pm t^\prime)$ is not at all linear. Most importantly, the curves for $t - t^\prime$ and $t + t^\prime$ are completely different, viz. $\langle \etaMD(t+t^\prime) \vert (\etaMD(t) = z) \rangle \neq \langle \etaMD(t) \vert (\eta(t+t^\prime) = z) \rangle$. This result therefore reveals explicitly that there is a very non-trivial irreversibility in the dynamics of $\etaMD.$ We will see more of it in Sec. \ref{sec:Dance}.

	\subsection{Broken fluctuation-dissipation relation}
	\label{sec:FDT}

    \begin{figure}
		\includegraphics[width=1.0\linewidth]{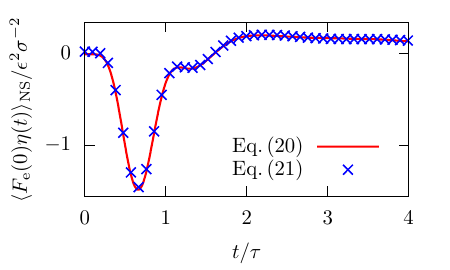}
		\caption{Non-stationary cross-correlation between external force and noise in system B0A, causing violation of the FDR. Direct measurement of the cross-correlations as extracted from Eq.~(\ref{eq:FDRB3}) (data points) compared to the violation of the FDR in Eq.~(\ref{eq:FDRB1}) (full line).}
		\label{fig:FigFDTB}
	\end{figure}
	
	\bj{The FDR is fulfilled if either the oscillator coupling or the external force are linear \cite{Zwanzig2001,JungJung2023}, i.e., in systems FH, FA, HA and BH (see also Appendix \ref{app:ValidFDR}).} \gj{It has been shown, however, that in fully non-linear systems the standard FDR is broken and instead the following equation holds \cite{Vroylandt2022, shea2024force}},
	\begin{equation}\label{eq:FDRB1}
		\langle F_\text{e}(0)\eta(t) \rangle_{\text{NS}} = \kB T K(t) - \langle \etaMD(0)\etaMD(t) \rangle.
	\end{equation}
	Here, the subscript NS indicates that the non-stationary cross-correlation is meant \bj{(cf. Appendix \ref{app:BrokenFDR})}, connected to Eq.~(\ref{eq:GLE}). This equation therefore states that emergent cross-correlations between external force and fluctuating force lead to additional contributions to the FDR. To connect this to the fluctuating force $\etaMD$, we derive a relation between the non-stationary correlation functions and the ones extracted from MD simulations (cf. Appendix \ref{app:BrokenFDR}), 
	\begin{align}\label{eq:FDRB3}
		\langle F_\text{e}(0)\eta(t) \rangle_{\text{NS}} &= \langle F_\text{e}(0)\etaMD(t) \rangle^\text{MD} \\& \quad - \int_{-\infty}^0 K_\text{V}(t-s) \langle v_0(s)F_\text{e}(0) \rangle^\text{MD} \diff s \nonumber.
	\end{align}
	Superscript MD indicates that these correlations are determined in MD simulations.
	
    Investigating the reconstructed fluctuating force $\etaMD$, we find indeed that for fully non-linear systems the non-stationary cross-correlations between external and fluctuating force are non-zero (data points in Fig.~\ref{fig:FigFDTB}). Even more importantly, by subtracting the autocorrelation of the noise from the memory kernel we can numerically confirm the non-trivial additional term in Eq.~(\ref{eq:FDRB1}) which enters the fluctuation-dissipation relation (lines in Fig.~\ref{fig:FigFDTB}). The dependence of $\langle F_\text{e}(0)\eta(t) \rangle_{\text{NS}}$ on the height of a barrier in the external potential is shown in Appendix \ref{app:FDRBBarr}.
    
    These results indicate that such cross-correlations need to be considered when creating fluctuating forces in non-linear GLE simulations if the microscopic fluctuations shall be modeled accurately. Since this has not been done in all previous applications of the non-linear GLE, it explains the observed deviations from the microscopic VACF \cite{Klipp2021,klippenstein2023bottom}.

	\subsection{Position dependence of noise and dissipative force}
	\label{sec:PosDep}

    The correlation with the external force, which depends on the position of the particle, of course poses the question whether the fluctuating force itself may also depend non-trivially on the position of the tagged particle \cite{plyukhin2026langevinpmf}. In the following, we therefore study the average fluctuating force, $\langle \etaMD \rangle(x_0^\prime) \equiv \langle \etaMD \rangle_{x_0^\prime} = \mathcal{T}^{-1} \int_0^\mathcal{T} \etaMD(t) \delta (x_0^\prime - x_0(t)) \difft $, along an MD trajectory of length $\mathcal{T}$, conditioned on the instantaneous position $x_0$ of the tagged particle.

	\begin{figure}
		\includegraphics[width=0.95\linewidth]{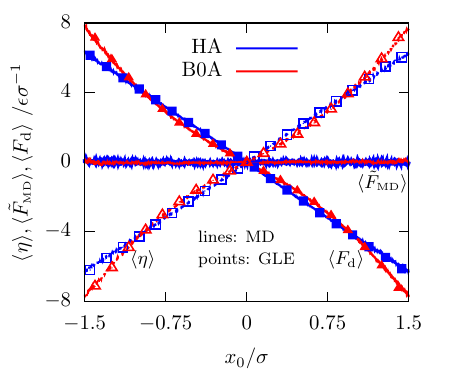}
		\caption{Position dependence of dissipative and fluctuating force. The conditional averages $\langle \eta \rangle_{x_0}$, $\langle F_{\text{d}} \rangle_{x_0}$ and $\langle \FMD \rangle_{x_0}$ are plotted as functions of the position of the tagged particle for systems with \bj{quadratic external potential} (HA, blue, squares) and \bj{quartic external potential} (B0A, red, triangles). Lines show results from MD, points from GLE simulations.}
		\label{fig:FigPos}
	\end{figure}
	
	We observe that the average value of the dissipative force $\langle F_{\text{d}} \rangle$, calculated according to the GLE (Eq.~(\ref{eq:GLE})), depends on the instantaneous position of the particle $x_0$ (see Fig.~\ref{fig:FigPos}). The steeper the external potential, the steeper the decay of $\langle F_{\text{d}} \rangle$ with $x_0$. For a linear external force, $\langle F_{\text{d}} \rangle (x_0)$ is a straight line with negative slope. For non-linear external forces, however, the position-dependence will become non-linear and can even become non-monotonic (see Appendix \ref{app:PosDepBarr}). The mean of $\FMD = F_{\text{d}} + \etaMD$, however, is zero for any position in all systems (horizontal lines in Fig.~\ref{fig:FigPos}), since it only depends on the relative distance between tagged particle and oscillator. In consequence, the noise shows the same position dependence as $F_{\text{d}}$, $\langle \eta \rangle (x_0) = - \langle F_{\text{d}} \rangle (x_0)$. These observations apply to $\langle \etaMD \rangle$ and $\langle \etaGLE \rangle$, i.e., to MD and GLE simulations (see Fig.~\ref{fig:FigPos}).
	
The reason for the emergence of this position dependence is the fact that the dissipative force $F_{\text{d}}$ does not only depend on the instantaneous velocity  of the particle as in the Langevin Eq.~(\ref{eq:LangEq}), but on its entire past trajectory. To rationalize this position dependence, we start with the dissipative force $F_{\text{d}}(t)$, which we can transform,
\begin{align}
	F_{\text{d}}(t) &= -\int_{-\infty}^t K(t-s) v_0(s) \diff s\\
	& = -\int_{-\infty}^t K(t-s) \frac{\diff}{\diff s} x_0(s) \diff s\\
	&= \int_{-\infty}^t x_0(s) \frac{\diff}{\diff s} K(t-s) \, \diff s \, - K(0) x_0(t), 
\end{align}
where we have used integration by parts in the second step. We can transform this further by changing the variables $t-s \rightarrow s$ in the integral,
\begin{align}
	F_{\text{d}}(t) &= - \int_{0}^\infty x_0(t-s) \frac{\diff}{\diff s} K(s) \, \diff s \, - K(0) x_0(t) \\
	&= K(0) \left(  \overline{ \mathfrak{T}[x_0(t)]} - x_0(t)  \right).
\end{align}
Here we have introduced the weighted average over the past trajectory of the particle, $\mathfrak{T}[x_0(t)]$, defined as,
\begin{align}
	\overline{ \mathfrak{T}[x_0(t)]} = - \frac{1}{K(0)} \int_{0}^\infty x_0(t-s) \frac{\diff}{\diff s} K(s) \, \diff s,
\end{align}
where the weight factor is the derivative of the kernel with respect to time. Finally, we calculate the conditional average $\langle ... \rangle_{x'_0}$ over all particle trajectories ending at a given $x'_0$,
\begin{equation}\label{eq:PosDep}
	\langle F_{\text{d}} \rangle (x'_0) \equiv \langle F_{\text{d}}(t) \rangle_{x'_0} = K(0) \left(  \langle \overline{\mathfrak{T}[x_0(t)]} \rangle _{x'_0} - x'_0  \right).
\end{equation}
We thus learn that the dissipative force is $K(0)$ times the difference between instantaneous position and a weighted average of the past trajectory, with weights given by the time-derivative of the memory kernel $\frac{\diff}{\diff t} K(t)$. Note that this whole derivation is valid for the GLE in general, independent of our model.

For example, for a linearly decaying memory kernel,
\begin{equation}
	K(t) = \begin{cases}
		K(0) - K(0) \frac{t}{t_\text{max}} & \text{for } t < t_\text{max}   \\
		0 & \text{otherwise}
	\end{cases}
\end{equation}
we find $\overline{ \mathfrak{T}[x_0(t)]} = \frac{1}{t_\text{max}} \int_{0}^{t_\text{max}} x_0(t-s) \diff s$, i.e., it equals the arithmetic mean of the particle's position on the memory time scale $t_{\text{max}}$. 

Another example: If the kernel decays exponentially, $K(t) = K(0) \text{e}^{- \frac{K(0)}{\gamma}t}$ where $\gamma$ is the integral over the kernel, then $\overline{ \mathfrak{T}[x_0(t)]} = - \frac{1}{\gamma} \int_{0}^\infty x_0(t-s) K(s) \diff s$, i.e., it is the average of the positions in the past trajectory, weighted with the kernel itself.

In consequence, in an external potential with a single minimum, the average will always be closer to this \emph{global} minimum than the current position, explaining the monotonic dependence of $\langle F_{\text{d}} \rangle$ on $x_0$ and the fact that the steeper the external potential, the steeper the decay of $\langle F_{\text{d}} \rangle (x_0)$. In a symmetric potential, these curves will pass the origin. In a double-well potential with sufficiently large barrier, the derivative of $\langle F_{\text{d}} \rangle (x_0) = K(0) \left(  \langle \overline{\mathfrak{T}[x_0(t)]} \rangle _{x_0} - x_0  \right)$ and hence of $\langle \eta \rangle (x_0)$ with respect to $x_0$ can easily change sign since the average will be influenced by \textit{both} neighbored \textit{local} minima, cf. Appendix \ref{app:PosDepBarr}.

\subsection{Irreversibility of dissipative force}
\label{sec:IrreversDiss}

\begin{figure}[h!]
	\includegraphics{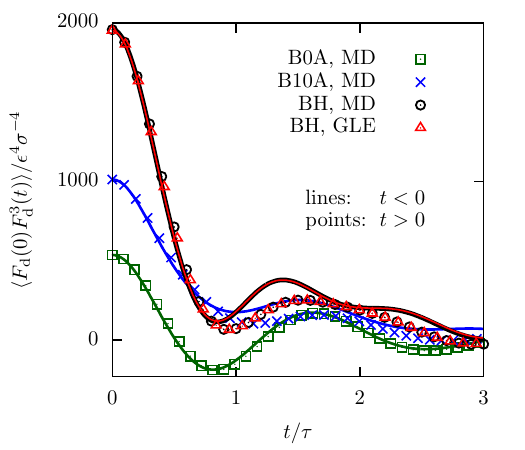}
	\caption{Irreversibility of the dissipative force $F_\text{d}(t)$. $\langle F_\text{d}(0)F_\text{d}^3(t) \rangle$ from MD simulations (system BH also GLE) for three systems. B0A: no barrier in $V_\text{e}$; B10A: high barrier; BH: like B10A but with harmonic oscillator coupling. Lines show results for $t<0$, points for $t>0$. }
	\label{fig:FigAsymFd}
\end{figure}

Knowing that $\etaMD(t)$ is not time-reversible, it stands to reason that $F_{\text{d}}(t)$ is not either and that these irreversibilities are analogous in some way. Fig.~\ref{fig:FigAsymFd} shows that irreversibility can indeed be observed for $F_{\text{d}}(t)$ as well. Importantly, this irreversibility is not connected to the trivial time-irreversibility due to the sign flip of velocities under time reversal, since the fourth-order correlation shown in the figure is blind to this property. What we observe is that the irreversibility of the fluctuating force $\etaMD$ and the dissipative force occur under very different conditions. \gj{$\etaMD(t)$ is irreversible as soon as the force $\FMD$ is anharmonic even in the absence of an external potential. In this regime $F_{\text{d}}(t)$ is reversible. This result may seem counterintuitive, since the total force $\tilde{F}(t) = F_\text{d}(t) + \etaMD(t)$ is inherently reversible. In Appendix \ref{app:IrrevPlusRev} we therefore rationalize why `reversible + irreversible = reversible' is possible. If we add a high barrier in the external potential, $F_\text{d}(t)$ becomes irreversible even if $\FMD$ is linear and thus $\etaMD$ is reversible.}

\gj{Another distinction is that the irreversibility of $F_\text{d}(t)$ in MD is correctly reproduced by GLE simulations  (see Fig.~\ref{fig:FigAsymFd}), in contrast to that of $\eta(t)$.}
	
\section{Iterative optimization of non-linear GLEs}
\label{sec:Iterative}

In practical simulations of coarse-grained soft-matter systems, the usage of the GLE is relevant to ensure that the time scale of the coarse-grained model matches the one of the microscopic simulations. It is therefore natural to study whether we can solve the above issues, in particular the emergence of cross-correlations between external and fluctuating force,  by finding an optimized memory kernel which leads to the correct coarse-grained dynamics despite using standard Gaussian noise $\etaGLE$ to model the microscopic fluctuating force $\etaMD$. This approach has been successfully tested in Refs.~\cite{klippenstein2023bottom, Klippensteinetal2024, Wolf2025} using an iterative optimization. This finding leads to the fundamental question: For arbitrary external forces $- V_{\text{e}}^\prime(x_0(t))$, can we always find a memory kernel $K_\text{iter}(t)$ such that the GLE in Eq.~(\ref{eq:eomMD2}) produces the correct VACF?

In the absence of external forces the result is known. Since the VACF is a correlation function and all correlation functions are Nevanlinna functions, we know that a decomposition of the type $\langle v_0(0) v_0(t) \rangle = - \int_0^t K(t-s) \langle v_0(0) v_0(s) \rangle \diff s$ exists and that the emergent kernel $K(t)$ is a correlation function \cite{franosch2026fundamental}. The latter property is essential since the kernel must be connected to the correlation function of the fluctuating force $\etaGLE$ via the FDR.

\begin{figure}
	\includegraphics{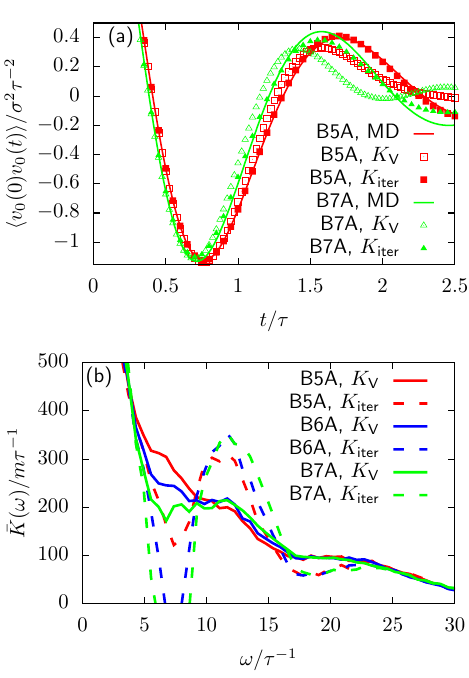}
    \caption{ Iterative optimization of the memory kernel to reproduce the correct microscopic VACF for different systems B$n$A. (a) VACF from MD and from GLE simulations using the Volterra kernel $K_\text{V}$ (Eq.~(\ref{eq:Kernel})) and the final iterative kernel $K_\text{iter}$ (see Appendix \ref{app:iterative}). (b) \gj{Real part of the} Fourier transform of the kernels $K_\text{V}(t)$ and $K_\text{iter}$ used in (a).   }
	\label{fig:iterative}
\end{figure}

In the presence of non-linear external forces the situation, however, is much less clear. To investigate this systematically, we study external potentials with barrier.  We introduce an iterative optimization of the memory similar to Ref.~\cite{JungHankeSchmid2017} but using an improved update scheme reminiscent of Ref.~\cite{klippenstein2023bottom}, which is detailed in Appendix \ref{app:iterative}.

We start with the system B5A, for which we observe a significant difference between the VACF observed in MD and GLE simulations using the Volterra kernel $K_\text{V}$ (see Fig.~\ref{fig:iterative}(a)). This difference can be much larger for highly non-linear soft-matter systems  such as the ones studied in Refs.~\cite{Klipp2021, klippenstein2023bottom,  Wolf2025}. This difference emerges since we have neglected the cross-correlations discussed in Sec.~\ref{sec:FDT}. After performing the iterative optimization, the GLE simulations with $K_\text{iter}(t)$ now perfectly reproduce the VACF. The situation, however, changes drastically for the system B7A. Even after many iterations, the GLE with $K_\text{iter}(t)$ does not converge to the MD results. 

This surprising observation is rationalized in Fig.~\ref{fig:iterative}(b), which shows the discrete Fourier transform of the memory kernel $\bar{K}(\omega) = \sum_t K(t) \cos(\omega t \Delta t)$. We observe that there are strong corrections in the optimization from $K_\text{V} \rightarrow K_\text{iter}$ for a frequency of $\omega \approx 6.$ The corresponding time scale $2 \pi / \omega \approx 1$ clearly coincides with the time scale at which the maxima of the cross-correlations in Fig.~\ref{fig:FigFDTBApp} are observed. However, when introducing higher barriers we find that these corrections grow in amplitude and finally lead to a clearly negative spectrum $\bar{K}(\omega).$ The transition seems to be located at the B6A system with barrier height $h = 0.75 \, \kB T. $ Above this transition, $K_\text{iter}(t)$ is not a correlation function anymore and it is impossible to create noise $\etaGLE$ fulfilling the FDR. Therefore, the iterative optimization does not converge. We therefore find that for systems with higher barriers ($> \kB T$) the iterative optimization fails. This observation has important consequences for coarse-grained soft-matter simulations, since it shows that regimes exist in which the ``standard'' non-linear GLE cannot be applied anymore. Concrete biomolecular system with significant barrier and non-Markovian dynamics are abundant (see, e.g., Refs.~\cite{dalton2023fast,Netz2025}). 
	
\section{The art of playback}
\label{sec:GLELPB}
	
	\subsection{How to create noise}
	\label{sec:HowNoise}
	
	We have already presented several differences between the noise $\etaMD$ calculated from MD (Eq.~(\ref{eq:MDnoise})) and the noise $\etaGLE$ usually used in GLE simulations (cf. Sec. \ref{sec:Methods}). \gj{In equilibrium systems, the latter is assumed to follow the FDR (Eq.~(\ref{eq:FDR})) and is usually modeled as a Gaussian distributed and time-reversible random process}, while the former may violate all these properties as shown in Sec.~\ref{sec:Analysis} \bj{(broken FDR and irreversibility) and, e.g., in Fig. 6 of Ref.~\cite{JungJung2023} (non-Gaussian $\etaMD$ in a system with anharmonic oscillator coupling)}.  In particular, the distribution of $\tilde{F}$, the sum of fluctuating and dissipative force, is strongly non-Gaussian in MD, but Gaussian in GLE simulations \cite{JungJung2023}. These discrepancies have consequences. Higher-order correlations cannot be reproduced correctly in GLE simulations if $\FMD$ is non-linear \cite{JungJung2023}, even when using the iterative optimization discussed in the previous section. These findings show that creating the correct noise for GLE simulations is both relevant and challenging \cite{widder2022generalized,zhu2023learning,JungHankeSchmid2018} and may require the usage of machine-learning techniques for more precise coarse-grained simulations, for example to consider complex cross-correlations.
    
    In the following, we want to take a complementary approach. Instead of creating new noise, we perform GLE simulations in which we recycle a time-series which actually fulfills all the properties we are interested in: the noise $\etaMD$ extracted from MD simulations (cf. Sec. \ref{sec:WhatIsNoise}).  We will call this method GLE playback (GLE-PB). In this chapter, we will study the consequences of using this MD noise as input for GLE simulations. In particular, we want to study whether it properly reproduces higher-order correlation functions, and whether there are any limitations for this approach. We also introduce the method `GLE-PB inv', which is identical to GLE-PB but uses a time-reversed fluctuating force to study the impact of the irreversible noise.
    
    GLE-PB corresponds to a method described recently by Kiefer et al. \cite{Netz2025}, which they call `non-Gaussian orthogonal force' GLE simulations. In our manuscript, we will focus on results that go beyond the analysis presented in their paper.
	
	In addition to GLE-PB, we will also analyze what happens if we do not calculate the dissipative force via the GLE convolution integral as in Eq.~(\ref{eq:MDnoise}) but via a simple Markovian expression as in the Langevin equation. The corresponding fluctuating force $\etaMDL(t)$ can be extracted from MD simulations,
	\begin{align}\label{eq:Lnoise}
		\etaMDL(t) &= F_0(t) - F_\text{e}(x(t)) - F_\text{d}(t)   \nonumber \\
        &= \FMD(t) - F_\text{d}(t)  \\
		F_\text{d}(t) &= - \gammaLPB v_0(t),   \nonumber
	\end{align}
	Using such noise in Langevin simulations, applying Eq.~(\ref{eq:LangEq}) with an arbitrary friction constant $\gammaLPB$ instead of $\gamma$, we will call Langevin playback (L-PB),
	\begin{equation}\label{eq:LangPB}
		F(t) = m \frac{\diff v}{\difft} = F_\text{e}(t) - \gammaLPB v(t) + \etaMDL(t).
	\end{equation}
	Therefore, we will distinguish now between $\etaMD$ determined via Eq.~(\ref{eq:MDnoise}) and $\etaMDL$ from Eq.~(\ref{eq:Lnoise}). Please note that while Eq.~(\ref{eq:LangPB}) may appear to be Markovian, it is in fact highly non-Markovian due to the complex correlations hidden in the fluctuating force $\etaMDL(t)$.

    For L-PB we used $\gammaLPB = 0.33 \gamma$, i.e., one third of the friction coefficient $\gamma = 15$ acting on the oscillator in our MD simulations of systems B$n$A. The reason was that in MD simulations $\gamma$ acts only on the oscillator but dampens the whole system with a weight of $m_0 + m_1 = 3$. To achieve comparable damping in L-PB, where $\gammaLPB$ acts directly on the particle with $m_0=1$, we chose this lower value of $\gammaLPB$. Generally, we found that $\gammaLPB = 0.33 \gamma$ was the lowest possible $\gammaLPB$ which could reproduce all relevant quantities in L-PB correctly. Larger friction, $\gammaLPB \gg 0.33 \gamma$, can always be used and usually leads to improved stability of L-PB.

	\subsection{Particles dancing to the tune of noise}
	\label{sec:Dance}

    	\begin{figure}
		\includegraphics[width=0.95\linewidth]{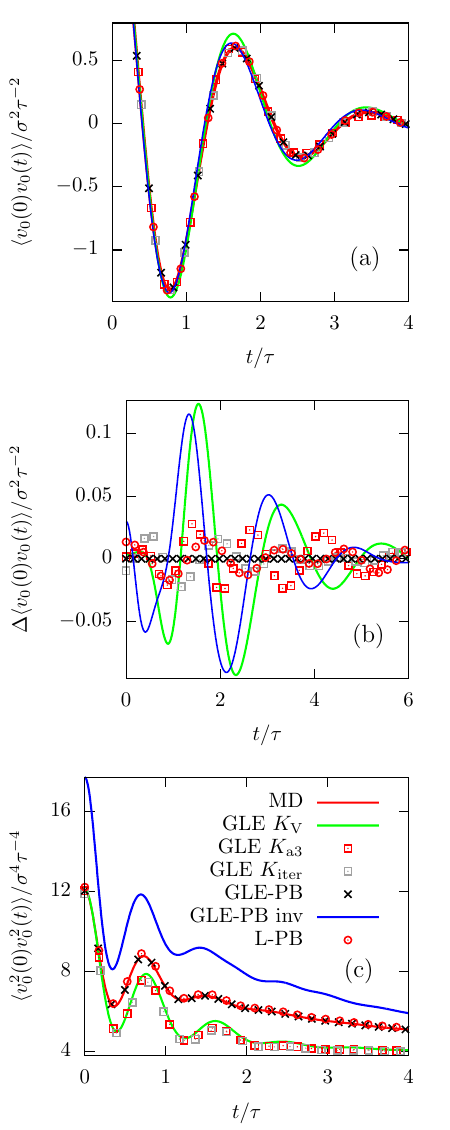}
		\caption{Reproduction of velocity correlations for the non-linear system B0A without barrier, using different simulation methods. (a) VACF, (b) deviation of VACF from the MD curve (i.e., full red line in (a)), (c) non-linear velocity correlations $\langle v_0^2(0)v_0^2(t) \rangle$. GLE-PB inv is GLE-PB with time-inverted $\etaMD$.}
		\label{fig:Fig_cv_39}
	\end{figure}
	
	We start studying non-linear systems without barrier. Consistent with the results discussed in Sec.~\ref{sec:Iterative}, we find for such systems that GLE simulations with $K_\text{iter}$ correctly reproduce the VACF of the MD simulations, while $K_\text{V}$ leads to significant deviations (see Fig.~\ref{fig:Fig_cv_39}(a),(b)). We also introduce an empirically corrected memory kernel $K_{\text{a}3}(t)$ which explicitly attempts to subtract the cross-correlations $\langle F_\text{e}(0)\eta(t) \rangle_\text{NS}$ from $K_\text{V}$  (see Appendix \ref{app:ModKern}).  We find that for systems without barrier the correction is very good. However, none of the above discussed GLE techniques is able to reproduce higher-order correlation functions which show very significant deviations from the MD results (see Fig.~\ref{fig:Fig_cv_39}(c) and Appendix \ref{app:ForceCorr}).

    These shortcomings are resolved when using GLE-PB or L-PB. We observe that both techniques not only reproduce the correct VACF but also precisely the higher-order correlation functions, which is of course very relevant to perform coarse-grained simulations with fully consistent dynamical properties (see Fig.~\ref{fig:Fig_cv_39}). In contrast to these results, when using a time-inverted fluctuating force in GLE-PB inv the dynamical properties of the coarse-grained simulations deviate strongly from MD. From our investigation we can therefore draw the important conclusion that the irreversibility of the fluctuating force matters and that it is \textit{not} sufficient to produce a noise in GLE simulations which has the correct probability density and the correct autocorrelation functions $\langle \etaGLE^k(0) \etaGLE^k(t) \rangle$ for any $k$, because all these properties are fulfilled by GLE-PB inv.

    It may be argued that the observed correct reproduction of the MD correlation functions is trivial, since the same noise is applied in GLE playback as was extracted from the microscopic MD simulations. However, consistent with Ref.~\cite{Netz2025}, we did not initialize GLE-PB with the same velocity history and the same initial conditions as the MD simulations. The extracted fluctuating force $\etaMD$ is therefore inconsistent with the external forces $F_\text{e}(t)$ calculated in GLE simulations. In consequence, it is actually very non-trivial why this inconsistency does not lead to inaccurate results. The first hint to resolve this paradox can be observed in Fig.~\ref{fig:Fig_cv_39}(b), which highlights that the difference between MD and GLE-PB is basically zero - much smaller than any statistical errors we would expect due to the limited length of the trajectories used to calculate the VACF. 

        \begin{figure}
    	\includegraphics[width=0.95\linewidth]{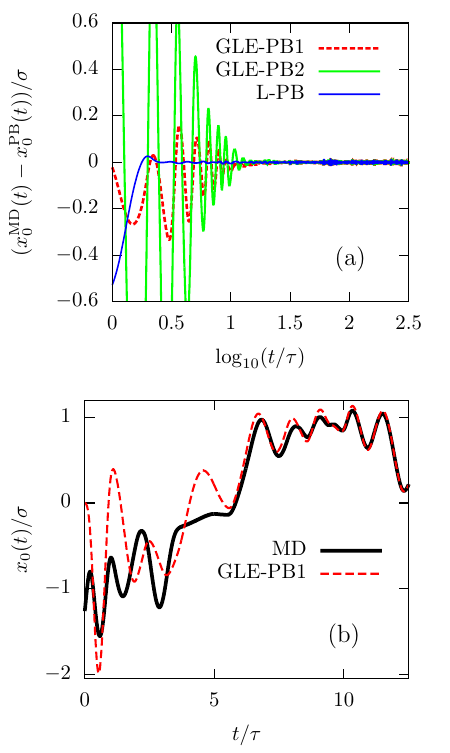}
    	\caption{Synchronization in GLE-PB and L-PB for system B0A. (a) Logarithmic plot of the position difference between PB and MD. For GLE-PB1 and GLE-PB2 different starting conditions have been used. (b) Trajectories of MD and GLE-PB1 in the beginning of synchronization.}
    	\label{fig:Fig_Posdiff39}
    \end{figure}
    
    To investigate the origin of this surprising result, we explicitly study the difference in the position of the particle between the original MD simulation and GLE playback in a system without barrier. We find that after a relatively short period of about $t \approx 10 \tau$ there is virtually no more difference between the positions in GLE-PB and MD (see Fig.~\ref{fig:Fig_Posdiff39}(b)). The trajectories are perfectly synchronized despite the completely different initial conditions. This observation holds independent of initial conditions, and L-PB reaches this state of synchronization even much faster  (see Fig.~\ref{fig:Fig_Posdiff39}(a)). The noise therefore clearly dictates the path of the particle in both playback methods, very different from the observations in Ref.~\cite{Netz2025}, where initially desynchronized trajectories remained desynchronized.

How could we explain this remarkable synchronization effect? Let us consider L-PB compared to the original MD simulation, and let us calculate the instantaneous deviation of total force $F_0$ in L-PB from MD as a consequence of a little deviation of the position $x_0$ in L-PB from MD. This $\Delta F_0(\Delta x_0)$ is the sum of changes in dissipative and external forces only, since the noise $\etaMDL$ is exactly the same in L-PB and MD. There is an implicit change of velocity in a single time step: $\Delta v_0 = \Delta x_0 / \Delta t$,
\begin{align}\label{eq:DeltaFsofort}
	\Delta F_0(\Delta x_0) &= \Delta F_\text{d}(\Delta v_0) + \Delta F_\text{e}(\Delta x_0)   \nonumber   \\
    &= -\gammaLPB \frac{\Delta x_0}{\Delta t} + \frac{\Delta F_\text{e}}{\Delta x_0} \Delta x_0   \nonumber   \\
	&\approx \left( -\frac{\gammaLPB}{\Delta t} + \frac{\diff F_\text{e}}{\diff x_0} \right) \Delta x_0   \\
	&= \left( -\frac{\gammaLPB}{\Delta t} - a_\text{e} - 3 b_\text{e} x_0^2 \right) \Delta x_0.   \nonumber 
\end{align}
In GLE-PB, $K_\text{V}(0)$ plays the role of $\frac{\gammaLPB}{\Delta t}$. As long as $a_\text{e} \ge 0$ (generally: as long as $\frac{\diff F_\text{e}}{\diff x_0} \le 0$), this $\Delta F_0$ definitely has the opposite sign to $\Delta x_0$ and thus acts to reduce $\Delta x_0$. This mechanism indicates how playback leads to synchronization.

This analysis is surely not a proof of dynamic stability since we have ignored several points. First of all, the fact that $\Delta F_0(\Delta x_0)$ acts opposite to $\Delta x_0$ is stabilizing but does not guarantee stability. Furthermore, there are not only changes in $x_0$ but also explicit deviations in $v_0$. And last but not least, in GLE-PB there will be also retarded effects of such deviations because of the memory. Anyway, a full Lyapunov stability analysis would go beyond the scope of this study, and Eq.~(\ref{eq:DeltaFsofort}) at least provides a qualitative explanation for the observed stable synchronization. It also gives a hint that negative $a_\text{e}$, i.e., a barrier in the external potential, might destabilize it.

\subsection{Desynchronization}
\label{sec:Desync}

\begin{figure}
    	\includegraphics[width=0.95\linewidth]{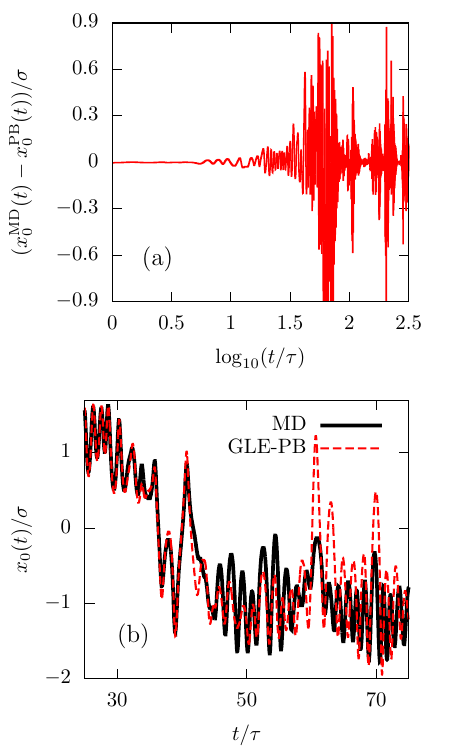}
    	\caption{Desynchronization in GLE-PB for system B9A. (a) Logarithmic plot of the position difference between GLE-PB and MD. The plot starts in a synchronized stage of the simulation. (b) Trajectories of MD and GLE-PB in the beginning of desynchronization.}
    	\label{fig:Fig_Posdiff39_9}
    \end{figure}
	
So far we showed that particles are dancing to the tune of noise in playback simulations, thus contrasting the results of Ref.~\cite{Netz2025}. In this reference a highly non-linear external potential with a high barrier ($> 3 \, \kB T$) has been investigated. Additionally, our stability analysis has indicated that the presence of a barrier may lead to destabilizing effects. Therefore, we now investigate the impact of GLE-PB on the system B9A which features a barrier of height $h \approx 1.7 \, \kB T$.

We clearly observe a qualitative difference to the systems without barrier studied in the previous section. It is evident that for a system with high barrier the particle position in GLE-PB desynchronizes from the MD simulation (see Fig.~\ref{fig:Fig_Posdiff39_9}(b)). Although the two trajectories started nearly perfectly synchronized, after some time they are strongly desynchronized. Thereafter, they stay asynchronous, interrupted by short intervals of synchronization (see Fig.~\ref{fig:Fig_Posdiff39_9}(a)). Which consequences has this on the reproduction of relevant quantities in GLE-PB?

	\begin{figure}[h!]
		\includegraphics[width=0.95\linewidth]{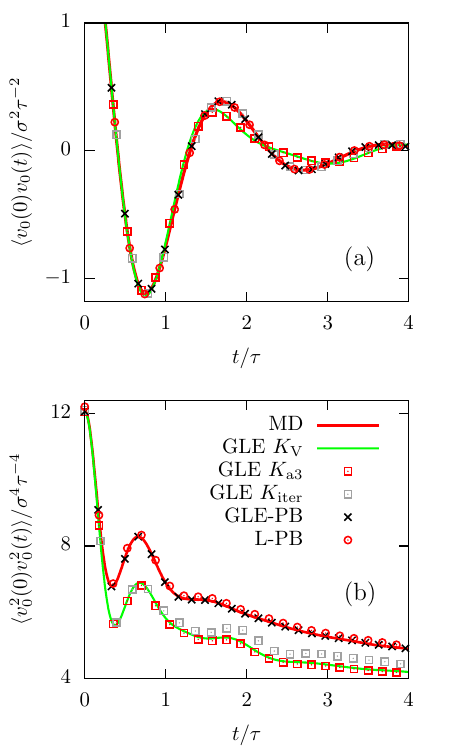}
		\caption{Dynamics in systems with low potential barrier. (a) VACF and (b) higher-order correlation $\langle v_0^2(0)v_0^2(t) \rangle$ for system B5A (barrier height $h=0.52 \, \kB T$).}
		\label{fig:Fig_cv_39_5}
	\end{figure}

    \begin{figure}
		\includegraphics[width=0.95\linewidth]{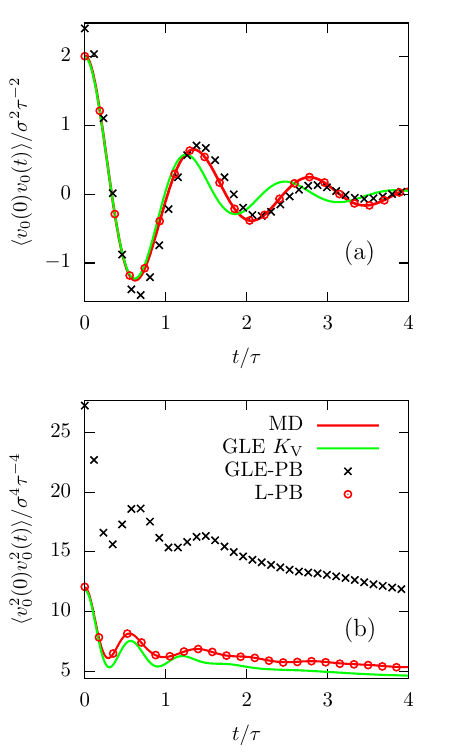}
		\caption{Dynamics in systems with high potential barrier. (a) VACF and (b) higher-order correlation  $\langle v_0^2(0)v_0^2(t) \rangle$ for system B10A (barrier height $h=2.08 \, \kB T$).}
		\label{fig:Fig_cv_39_10}
	\end{figure}

We therefore investigate systematically what happens when the barrier is increased, starting from a system with low barrier ($h \approx 0.5 \, \kB T$). \bj{To avoid any sampling issue, simulations were run including at least $10\,000 - 100\,000$ barrier crossings, depending on barrier height.} We find that the VACF is reproduced correctly by both PB methods and also by GLE with an iteratively optimized kernel ($K_\text{iter}$). $K_\text{V}$ and the kernel modified via Eq.~(\ref{eq:Ka3}) ($K_\text{a3}$), however, fail (see Fig.~\ref{fig:Fig_cv_39_5}). The autocorrelation of $v_0^2$, i.e., a higher-order correlation, is described correctly by both PBs, but not by the other techniques. Apart from the failure of $K_\text{a3}$ this system therefore behaves similarly to the one without barrier. If we examine a system with high barrier ($h \approx 2 \, \kB T$, Fig.~\ref{fig:Fig_cv_39_10}), however, only L-PB reproduces VACF and $\langle v_0^2(0)v_0^2(t) \rangle$ correctly, while GLE-PB even fails to produce the proper values of $\langle v_0^2 \rangle$ and $\langle v_0^4 \rangle$. 

\begin{figure}
	\includegraphics[width=0.95\linewidth]{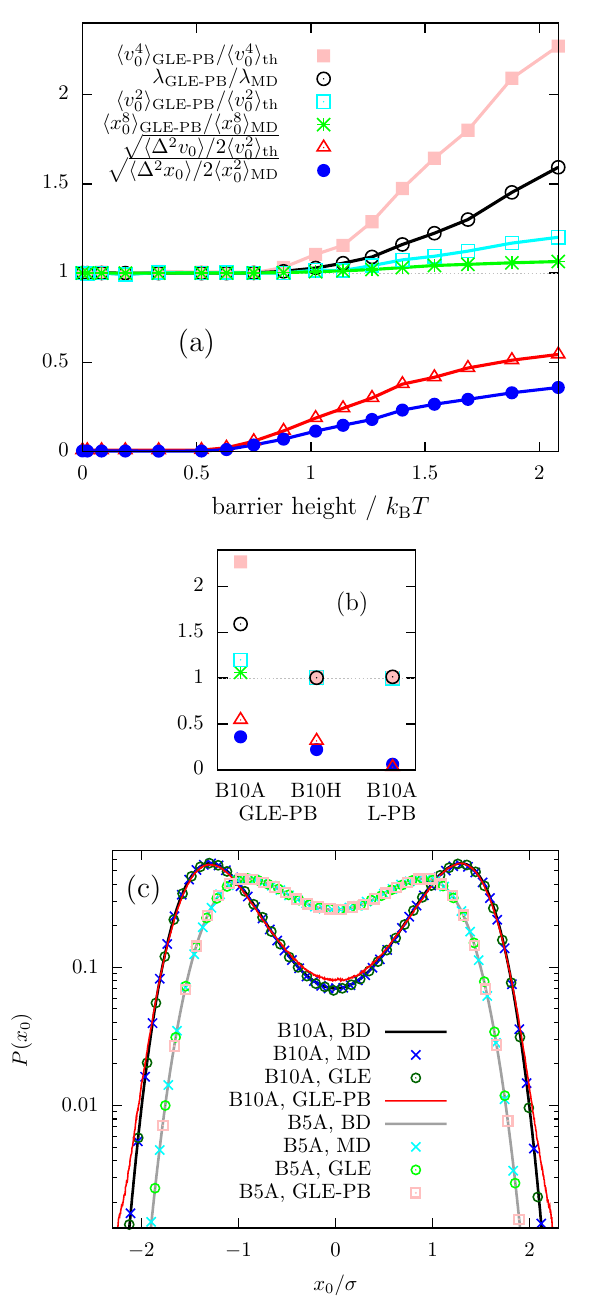}
	\caption{Development of desynchronization in GLE-PB. (a) Different ratios and relative deviations for systems B$n$A, plotted vs. barrier height $h$. Subscript 'th' indicates the theoretical value according to the Boltzmann distribution. $\lambda$ is the frequency of barrier crossing. $\langle \Delta ^2 x_0 \rangle$ and $\langle \Delta ^2 v_0 \rangle$ are the mean-square differences between GLE-PB and MD of the particle's position and velocity, respectively. (b) Comparison of the values for the highest barrier from (a) (B10A, $h = 2.08 \, \kB T$) with GLE-PB results of system B10H (the same external potential but harmonic oscillator coupling). The third column shows the L-PB results for system B10A. \bj{(c) Logarithmic plots of the normalized probability distribution of $x_0$ for systems with high (B10A) and low barrier (B5A) from different simulation methods vs. the curves expected from the Boltzmann distribution (BD). }}
	\label{fig:Fig_GLEPB_Abw}
\end{figure}

These results show that desynchronization has fatal consequences for GLE-PB. In Fig.~\ref{fig:Fig_GLEPB_Abw}(a) we analyze the development of various dynamical properties as a function of the barrier height. The first deviations of position and velocity in GLE-PB from MD show up above $0.5 \, \kB T$. Relevant values only begin to differ at a slightly higher barrier. Obviously there is a certain tolerance, depending on the quantity. Deviations of the VACF and of $\langle v_0^2 \rangle$ start relatively late and stay relatively small. Noteworthy are also the deviations in the frequency of barrier crossing $\lambda$. For barrier heights above $\kB T$, GLE-PB clearly fails to reproduce this important quantity, performing even worse than GLE, in contrast to the results shown in Ref.~\cite{Netz2025}. Strong and early deviations are found for the autocorrelation of $v_0^2$ and for $\langle v_0^4 \rangle$. The latter can serve as an indicator: if $\langle v_0^4 \rangle = 3 (\kB T/m_0)^2$ is found, all the other quantities should be correct, too. Interestingly, this synchronization/desynchronization transition occurs at exactly the same barrier height as the transition observed in the iterative optimization of the memory kernel in Sec.~\ref{sec:Iterative}. And it is the same barrier height at which the peak of the non-stationary cross-correlation of external force and noise shifts from negative to positive, cf. Appendix \ref{app:FDRBBarr}.

The most significant failure of GLE-PB starts to become apparent at $h \gtrsim 2 \, \kB T$, where we observe that not only the velocity correlations but also the position correlations are not exactly reproduced, as shown by the \bj{8th-order moments in Fig.~\ref{fig:Fig_GLEPB_Abw}(a) and explicitly in Fig.~\ref{fig:Fig_GLEPB_Abw}(c). We observe that the Boltzmann distribution $P(x_0) \propto \exp(-V_\text{e}(x_0)/\kB T)$ is correctly reproduced by MD and GLE simulations. The only exception are GLE-PB simulations for the system with high barrier, which shows small but significant differences to the Boltzmann distribution. This implies that GLE-PB simulations with high barrier do not sample an equilibrium ensemble anymore.} This observation disqualifies GLE-PB as a suitable simulation method in the presence of high barriers with strong anharmonicities in the system.

How does this synchronization/desynchronization transition depend on the other parameters in the system? First, we find that when using L-PB even at high barrier the trajectories remain very well synchronized (see B10A L-PB in Fig.~\ref{fig:Fig_GLEPB_Abw}(b)). Therefore, the above findings depend sensitively on which coarse-grained model is used in playback simulations, consistent with conclusions in Ref.~\cite{Netz2025}. Second, when we choose the extreme case of a harmonic coupling between the tagged particle and the oscillator, all dynamical quantities are well reproduced, while there is no synchronization  (see B10H in Fig.~\ref{fig:Fig_GLEPB_Abw}(b)).  The two transitions observed in Fig.~\ref{fig:Fig_GLEPB_Abw}(a) are therefore not necessarily coupled, albeit they seem to occur at very similar barrier height in our systems. This means that, on the one hand, GLE-PB fails in reproducing the MD trajectories, but on the other hand, it succeeds as a normal GLE simulation in which the noise is produced independent of the current position and velocity and the goal is to reproduce the general dynamic properties of the system. Different from what we have observed for the dynamical properties in Ref.~\cite{JungJung2023} which depend strong on $\FMD$, synchronization is independent of the non-linearity of $\FMD$ but seems to depend on the barrier in the external potential only, i.e., desynchronization itself in GLE-PB might be connected to the irreversibility of $F_\text{d}(t)$ (cf. Sec. \ref{sec:IrreversDiss}) but \textit{not} to the irreversibility of $\etaMD(t)$ (Sec. \ref{sec:IrreversNoise}) and \textit{not} to the broken FDR (Sec. \ref{sec:FDT}). These results may indicate that desynchronization is linked to the well-known fact that Duffing oscillators tend to exhibit chaotic behavior \cite{LenciRega2011}. In Appendix \ref{app:DepInst}, we show in more detail how desynchronization in GLE-PB depends on parameters other than barrier height.
	
So far, we have considered only symmetric external potentials. Systems with asymmetric $V_{\text{e}}$ are discussed in Appendix \ref{app:AsymPot}. There it is shown that high shoulders in the external potential have essentially the same effect on GLE-PB as high barriers, as is to be expected from Eq.~(\ref{eq:DeltaFsofort}).

\subsection{Response to perturbation}
\label{sec:Perturb}

We have studied extensively the equilibrium dynamical properties of the system. However, it is similarly important to investigate transport properties, as measured by response functions in the linear regime. In this section we will therefore investigate the average velocity response of MD, GLE and playback methods to single, small perturbations. This is conducted with the following procedure: We run a simulation (MD, GLE or playback). At some time $t'$ we start a replica of this original simulation in exactly the same state using the same memory, except that in the replica a deviation of $\Delta v = 0.1 \, \sigma \tau^{-1}$ is added to the current velocity of the original. Now both simulations run in parallel with the same noise, and the deviation of velocity between original (superscript 'EQ') and replica ('R') is recorded. From numerous runs, started at different $t'$, the average is calculated  \cite{jung2021fluctuation},
\begin{align}\label{eq:Rt}
	R(t)  = \Bigl \langle v_0^\text{R}(t'+t) - v_0^\text{EQ}(t'+t) \Bigr \rangle / \Delta v.
\end{align}

\begin{figure}
	\includegraphics[width=0.95\linewidth]{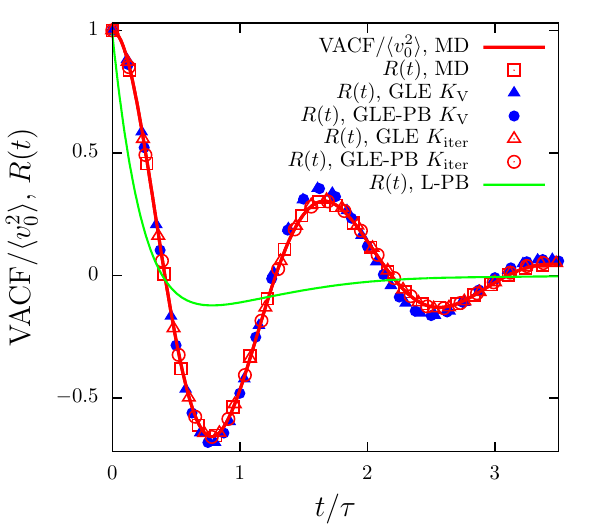}
	\caption{Response to perturbation $R(t)$ in MD, GLE and GLE-PB simulations with $K_\text{V}$ and $K_\text{iter}$ as well as in L-PB, compared to the normalized VACF from MD, for the fully anharmonic system B0A.}
	\label{fig:Fig_React}
\end{figure}

We find that the response $R(t)$ in MD is identical to the normalized VACF from MD, which is another manifestation of the fluctuation-dissipation relation \cite{kubo1966fluctuation,franosch2026fundamental}. Similarly, we observe that any GLE simulation using the iteratively optimized kernel $K_\text{iter}$ (or $K_\text{a3}$, not shown here) leads to the correct response function (see Fig.~\ref{fig:Fig_React}). This highlights another important aspect of the results shown in Sec.~\ref{sec:Iterative}. In contrast, any GLE using the kernel $K_\text{V}$, even GLE-PB, leads to an incorrect response. The most extreme case is, of course, L-PB which fails completely in predicting the correct response function $R(t)$. This shows that playback may be usable in reproducing the correct equilibrium dynamical properties, however, for the response functions the correct non-Markovian memory kernel $K(t)$ is required.
	
	\section{When can we use non-linear GLEs?}
	\label{sec:Whatwhere}
	
	Finally, we wrap up the results presented in this manuscript and show a very condensed summary of our findings when there is a strong anharmonic coupling between the tagged particle and the bath (see Tab.~\ref{tab:waswo}). GLE simulations using the Volterra kernel generally work perfectly whenever $\FMD$ is linear. This is, of course, the case for a harmonically coupled oscillator, but it also holds true when $\FMD$ is the sum of a sufficiently large number of sufficiently uncorrelated non-linear forces \cite{JungJung2023}.

	To summarize our results, standard GLE simulations consistently fail to reproduce higher-order correlations. In the presence of non-linear external or conservative forces, iterative optimization or $K_\text{a3}$ is required to even reproduce the VACF. In systems with high barrier, GLE completely fails to replicate dynamic parameters and only captures static properties. While GLE-PB improves the performance for low barriers, in particular with respect to higher-order correlations, high barriers lead to desynchronization and a complete failure in all properties, even static ones. Finally, L-PB always reproduces correctly the trajectories in all systems, but never the response to perturbations and thus the transport properties.

     \newcommand{\sqcell}{\rule{0.5cm}{0.0cm}}
	\begin{table}[htbp]
		\centering
		\begin{tabular}{|c||c|c|c||c|c|c||c|c|c|}
			\hline
			& \multicolumn{9}{|c|}{External force}  \\
			\hline
			& \multicolumn{3}{|c|}{linear} &  \multicolumn{6}{|c|}{non-linear}  \\
			& \multicolumn{3}{|c|}{ } & \multicolumn{3}{|c|}{low barrier} & \multicolumn{3}{|c|}{high barrier}  \\
            \hline
            \hline
			  & \tiny{GLE} & \tiny{G-PB} & \tiny{L-PB} & \tiny{GLE} & \tiny{G-PB} & \tiny{L-PB} & \tiny{GLE} & \tiny{G-PB} & \tiny{L-PB} \\
			\hline
			\hline
			VACF & \cellcolor{green}\sqcell & \cellcolor{green}\sqcell & \cellcolor{green}\sqcell & \cellcolor{green}\centering{$*$} & \cellcolor{green} & \cellcolor{green} & \cellcolor{red} & \cellcolor{red} \sqcell & \cellcolor{green} \\
			\hline
			$\langle v_0^2(0)v_0^2(t) \rangle$ & \cellcolor{red} & \cellcolor{green} & \cellcolor{green} & \sqcell \cellcolor{red} & \sqcell \cellcolor{green}  & \sqcell \cellcolor{green} & \cellcolor{red} \sqcell & \cellcolor{red} \sqcell & \cellcolor{green} \sqcell \\
			\hline
			$\langle F_\text{e}(0)F_\text{e}(t) \rangle$ & \cellcolor{green} & \cellcolor{green} & \cellcolor{green} & \cellcolor{red} & \cellcolor{green}  & \cellcolor{green} & \cellcolor{red} & \cellcolor{red} & \cellcolor{green} \\
			\hline
			$P(x_0)$ & \cellcolor{green} & \cellcolor{green} & \cellcolor{green} & \cellcolor{green} & \cellcolor{green}  & \cellcolor{green} & \cellcolor{green} & \cellcolor{red} & \cellcolor{green} \\
			\hline
			$\lambda$ & \centering{\bf{---}} & \centering{\bf{---}} & \centering{\bf{---}} & \cellcolor{green} & \cellcolor{green} & \cellcolor{green} & \cellcolor{red} & \cellcolor{red} & \cellcolor{green} \\
			\hline
			$R(t)$ & \cellcolor{green} & \cellcolor{green} & \cellcolor{red} & \cellcolor{green}\centering{$*$} & \cellcolor{green}\centering{$*$} & \cellcolor{red} & \cellcolor{red} & \cellcolor{red} & \cellcolor{red} \\
			\hline
		\end{tabular}
        		\caption{Which simulation reproduces which quantity correctly in which external potential (green: correctly reproduced, red: incorrect)? Here, $P(x_0)$ is the positional distribution function of the particle, $\lambda$ is the frequency of barrier crossing, and $R(t)$ is the response to a single perturbation. GLE uses standard correlated Gaussian noise, G-PB is short for GLE-PB. $K_\text{V}$ was used to calculate noise and dissipative force. Exception: an asterisk ($*$) indicates that this method only works correctly if the kernel is modified to correctly reproduce the VACF using, e.g., iterative optimization (Sec.~\ref{sec:Iterative}). 'Barrier' stands for 'barrier or shoulder'. 'High barrier' means that it is so high that GLE-PB with the Volterra kernel does not correctly reproduce $\langle v_0^4 \rangle$ (i.e., $h > \kB T$ in our model).}
		\label{tab:waswo}
	\end{table}

	\section{Conclusion}
	\label{sec:Conclusion}
	
\gj{Non-linear external or conservative forces are used frequently} in combination with non-Markovian memory kernels and correlated fluctuating forces to model soft-matter systems using generalized Langevin equations. In this work, we have shown that the usual approach of modeling these fluctuating forces as colored, Gaussian noise implies many approximations which have not yet been discussed in the literature. Among others, we have shown that the microscopic fluctuating forces are irreversible and have a non-trivial position dependence. Neglecting these properties can imply that all dynamical properties of the underlying microscopic system are not properly reproduced by the GLE.

Importantly, we have found a very pronounced dynamical transition between two types of non-linear external or conservative forces. In systems with a low barrier it is possible to iteratively optimize the memory kernel to correct for some of the inconsistencies in the dynamics emerging due to the non-linear forces, consistent with the results in Ref.~\cite{Klipp2021}. However, if there are high barriers or shoulders in the external potential, such an optimization is not possible anymore.

Finally, we have studied the consequences of performing coarse-grained simulations using fluctuating forces which were extracted from MD simulations (`playback') \cite{Netz2025}. We found that for low barriers, or when using Langevin friction, playback correctly reproduces all relevant higher-order correlation functions. Even more remarkably, playback perfectly synchronizes with the original MD simulations. As a consequence, the technique is not able to generate new trajectories which is a core requirement for coarse-graining procedures. For high barriers, GLE-PB desynchronizes from the MD trajectories, while it also leads to large deviations from the original dynamics and does not even sample from an equilibrium ensemble.  The systems analyzed in Ref.~\cite{Netz2025} therefore seem to represent a sweet spot, where playback does not lead to strong synchronization, while still improving the standard GLE results using colored, Gaussian noise. When increasing the number of oscillators $N>1$ we find also for our model specific regimes without synchronization where playback performs better than standard GLE (data not shown).  Our analysis therefore indicates that the observations in Ref.~\cite{Netz2025} do not represent the general case, and that such techniques need to be applied with great care. \gj{In future work, it would therefore be very important to conduct a similar systematic analysis as performed in this manuscript for more complex MD systems, including liquids \cite{klippenstein2023bottom}, polymers \cite{Netz2025,li2017computing} or proteins \cite{hery2026comparison}.}

Our work further highlights the importance of the fluctuating force in performing accurate coarse-grained simulations in the presence of non-linear forces. A new simulation method is needed, and we are convinced that this will not be possible without \gj{position-dependent} or multiplicative noise. \gj{Future directions could therefore include the application of} generative networks which can be machine-learned to produce fluctuating forces with all the dynamical properties of the underlying microscopic fluctuations. Such techniques may enable performing improved coarse-grained simulations of soft matter, in particular, biomolecular systems \cite{zhu2023learning,cherchi2024ml,ge2024data}. Additionally, it would be interesting to expand our analysis to multi-dimensional \cite{shea2024force,kiefer2025influence,bockius2026determining,hery2026non}, overdamped \cite{ma2016generalized} or non-equilibrium \cite{te2019mori,mitterwallner2020non,jung2022non,jung2024mobility,hery2024derivation} GLE models, which are all relevant for future applications. Finally, future direction could also include the analysis of non-linear memory \cite{vroylandt2022position,ayaz2022generalized,SciPostPhys.13.4.078}, for example using Markovian embedding \cite{jaganathan2025markovian}.

	\textit{Acknowledgments:} The authors would like to thank Viktor Klippenstein and Niklas Wolf for insightful discussions. This research was funded by the Austrian Science Fund (FWF) 10.55776/PAT1139125 (GJ).
	
	\, \\
	
	\appendix
	
	\begin{center}
		{\bf APPENDICES}
	\end{center}

\section{Derivation of Eqs.~(\ref{eq:condmean}) and (\ref{eq:condvar})}
\label{app:DerCond}

Let's assume that $\langle \eta(t) \vert (\eta(t-t^\prime) = \eta') \rangle = \mathfrak{a} \, \eta'$, i.e., that the average relative decay of the autocorrelation of noise $\eta$ over a time interval $t^\prime$ is the same for all values of $\eta'$. From this linear ansatz, we can calculate the slope $\mathfrak{a}$. $p$ is the probability density distribution of $\eta$.
\begin{align}\label{eq:mathfrak_a1}
	&\langle \eta(t)\eta(t-t^\prime) \vert (\eta(t-t^\prime) = \eta') \rangle =   \nonumber \\
    &\langle \eta(t) \vert (\eta(t-t^\prime) = \eta') \rangle \, \eta'
\end{align}
\begin{align}\label{eq:mathfrak_a2}
	\langle \eta(t)\eta(t-t^\prime) \rangle &= \int_{-\infty}^{\infty} p(\eta') \, \langle \eta(t) \vert (\eta(t-t^\prime) = \eta') \rangle \, \eta' \, \diff \eta'   \nonumber \\
	&= \int_{-\infty}^{\infty} p(\eta') \, \mathfrak{a} \, \eta'^2 \, \diff \eta' = \mathfrak{a} \langle \eta^2 \rangle
\end{align}
Therefore, we find that
\begin{align}\label{eq:mathfrak_a3}
	\mathfrak{a} = \frac{\langle \eta(t)\eta(t-t^\prime) \rangle}{\langle \eta^2 \rangle} = \frac{K_{\text{V}}(t^\prime)}{K_{\text{V}}(0)}
\end{align}

The conditional variance of noise is
\begin{align}\label{eq:mathfrak_c1} 
	\sigma^2_{\eta(t) \vert (\eta(t-t^\prime) = \eta')} &= \langle \eta^2 \rangle_{t^\prime,\eta'} - \langle \eta(t) \vert (\eta(t-t^\prime) = \eta') \rangle^2   \nonumber \\
    &= \langle \eta^2 \rangle_{t^\prime,\eta'} - \mathfrak{a}^2 \eta'^2 = \mathfrak{c}^2
\end{align}
where $\langle \eta^2 \rangle_{t^\prime,\eta'} \equiv \langle \eta^2(t) \vert (\eta(t-t^\prime) = \eta') \rangle$.
Thus $\langle \eta^2 \rangle_{t^\prime,\eta'} = \mathfrak{a}^2 \eta'^2 + \mathfrak{c}^2$. Now we can calculate $\langle \eta^2 \rangle$ as a weighted average of all $\langle \eta^2 \rangle_{t^\prime,\eta'}$:
\begin{align}\label{eq:mathfrak_c2}
	\langle \eta^2 \rangle &= \int_{-\infty}^{\infty} p(\eta') \, \langle \eta^2 \rangle_{t^\prime,\eta'} \, \diff \eta' =  \int_{-\infty}^{\infty} p(\eta') \, (\mathfrak{a}^2 \eta'^2 + \mathfrak{c}^2) \, \diff \eta'  \nonumber \\
	&= \mathfrak{a}^2 \int_{-\infty}^{\infty} p(\eta') \, \eta'^2 \, \diff \eta' \, + \mathfrak{c}^2 = \mathfrak{a}^2 \langle \eta^2 \rangle + \mathfrak{c}^2 
\end{align}
\begin{align}\label{eq:mathfrak_c3}
	\mathfrak{c}^2 &= \langle \eta^2 \rangle (1 - \mathfrak{a}^2) =  \kB T K_{\text{V}}(0) \left( 1 - \frac{K_{\text{V}}^2(t^\prime)}{K_{\text{V}}^2(0)} \right)   \nonumber \\
	&= \frac{\kB T}{K_{\text{V}}(0)} (K_{\text{V}}^2(0) - K_{\text{V}}^2(t^\prime))
\end{align}

\begin{figure}
	\includegraphics[width=0.95\linewidth]{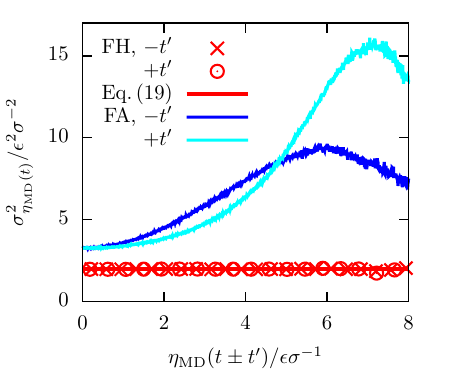}
	\caption{Conditional variance of $\etaMD(t)$ as a function of $\etaMD(t \pm t^\prime)$ ($t^\prime = 0.455 \tau$) for the same systems as in Fig.~\ref{fig:Figetalag}. The horizontal red line is calculated according to Eq.~(\ref{eq:condvar}).}
	\label{fig:Figsigmaetalag}
\end{figure}

We found that these relations always hold for $\etaGLE$, but for $\etaMD$ only if the oscillator coupling is linear, cf. Figs.~\ref{fig:Figetalag} and \ref{fig:Figsigmaetalag}.

\section{Broken FDR}

\subsection{Validity of the FDR for not fully anharmonic systems}
\label{app:ValidFDR}

If a system is not completely non-linear, i.e., if either external force \textit{or} oscillator coupling are linear, the FDR is valid. Fig.~\ref{fig:FigFDTBApp}(a) shows for three examples that $\kB T K(t)$ equals $\langle \etaMD(0)\etaMD(t) \rangle$ in such systems. This also indicates that the irreversibility of the noise is not linked to the violation of the FDR, because in systems FA and HA $\etaMD$ is irreversible, but the FDR is fulfilled.
\color{black}

\subsection{Derivation of Eq.~(\ref{eq:FDRB3}) and discussion of Eq.~(\ref{eq:FDRB1})}
\label{app:BrokenFDR}
	
    Eq.~(\ref{eq:FDRB1}) has been shown in Ref.~\cite{Vroylandt2022} for the non-stationary GLE where the convolution integral in the GLE starts at $t=0.$ However, when performing coarse-grained simulations in stationary systems, which is basically always the case, we want to start the convolution integral at $t \rightarrow - \infty.$ Here we show how we can still connect our results to the calculations in Ref.~\cite{Vroylandt2022}.

    We start with the general GLE,
	\begin{align}
		m_0 \frac{\diff v_0(t)}{\diff t} = F_\text{e}(t) - \int_{\text{T}}^t K(t-s) v_0(s) \diff s + \eta(t)
	\end{align}
    T $=0$ would correspond to the non-stationary GLE (NS) and T $=-\infty$ to the stationary GLE (S) used to extract the fluctuating force from our MD simulations. We multiply this equation with $x_0^n$ and take the average value,
	\begin{align}
		-m_0 \Bigl \langle x_0^n(0) &\frac{\diff v_0(t)}{\diff t} \Bigr \rangle = \langle x_0^n(0)F_\text{e}(t) \rangle + \langle x_0^n(0)\eta(t) \rangle \nonumber\\
        &- \int_{\text{T}}^t K(t-s) \langle v_0(s)x^n_0(0) \rangle \diff s 
	\end{align}
    We know that all correlation functions not involving the fluctuating force $\eta(t)$ are identical between the stationary and the non-stationary GLE. Therefore, we find
	\begin{align}\label{eq:cetax0_statminusnichtstat}
		\langle x^n_0(0)\eta(t) \rangle_\text{S} &- \langle x^n_0(0)\eta(t) \rangle_\text{NS} =   \nonumber \\
        &\int_{-\infty}^0 K(t-s) \langle v_0(s)x^n_0(0) \rangle \diff s
	\end{align}
	Assuming now that $F_\text{e}(x_0)$ is well-behaved and can be expanded in polynomials, we obtain 
	\begin{align}\label{eq:cetaFe_statminusnichtstat}
		\langle F_\text{e}(0)\eta(t) \rangle_\text{S} &- \langle F_\text{e}(0)\eta(t) \rangle_\text{NS} =   \nonumber \\
        &\int_{-\infty}^0 K(t-s) \langle v_0(s)F_\text{e}(0) \rangle \diff s
	\end{align}
    This equation allows us to transfer any results for the cross-correlations $\langle F_\text{e}(0)\eta(t) \rangle_\text{NS}$ derived for non-stationary systems to stationary GLE simulations, cf. Eq.~(\ref{eq:FDRB3}).

It is interesting that a relation very similar to Eq.~(\ref{eq:FDRB1}) was published for GLEs derived from stochastic microscopic dynamics in Ref.~\cite{Zhu_2023} (see their Eq.~(4)). In this reference, the additional term in the FDR emerges because the evolution operator is not self-adjoint. It appears that Eq.~(\ref{eq:FDRB1}) in the present manuscript is a special case of Eq.~(4) in Ref.~\cite{Zhu_2023} with $w(0) = F_\text{e}(0)$.

\subsection{Growing barrier heights}
\label{app:FDRBBarr}

\begin{figure}
	\includegraphics[width=0.95\linewidth]{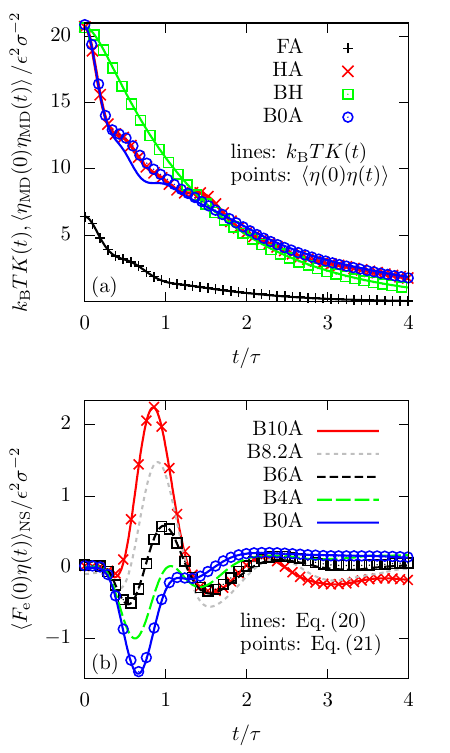}
	\caption{\bj{(a) Validity of the FDR in systems with anharmonic oscillator coupling but without (FA) or with harmonic external potential (HA), or with high barrier but harmonic oscillator coupling (BH). In the fully anharmonic system B0A, the FDR is broken. (b)} Non-stationary cross-correlation between external force and noise in systems with different barrier heights.}
	\label{fig:FigFDTBApp}
\end{figure}

In systems without a barrier in the external potential, $\langle F_{\text{e}}(0)\eta(t) \rangle_{\text{NS}}$ exhibits a dominant negative peak. However, when a barrier is introduced, a small positive peak develops next to the negative one. As the barrier increases, the negative peak shrinks and the positive grows. At a barrier height $h = 0.75 \, \kB T$, they exhibit approximately the same amplitude. While the barrier increases further, the positive peak becomes dominant, cf. Fig.~\ref{fig:FigFDTBApp}(b).
	
\section{Position dependence of noise in an external double-well potential}
\label{app:PosDepBarr}

\begin{figure}
		\includegraphics[width=0.95\linewidth]{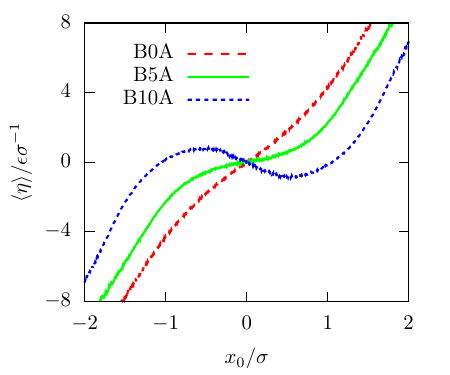}
		\caption{Position dependence of the conditional fluctuating force in external potentials with different barrier heights (cf. Fig.~\ref{fig:Figetalag}). B0A: no barrier, B5A: low barrier ($h \approx 0.5 \, \kB T$), B10A: high barrier ($h \approx 2 \, \kB T$).}
		\label{fig:FigPosBarr}
	\end{figure}

\gj{We have investigated in Sec.~\ref{sec:PosDep} the position-dependence of noise in convex potentials. In consequence, the position-dependence was always monotonic. In contrast,  Fig.~\ref{fig:FigPosBarr} shows that in a double-well potential the dependence of noise on position can become non-monotonic with growing barrier height.}

\section{Irreversible plus reversible is not necessarily irreversible}
\label{app:IrrevPlusRev}

We have reported in Sec.~\ref{sec:Analysis} that for specific parameters it is possible that $\tilde{F}(t) = \eta(t) + F_\text{d}(t)$ is reversible while noise $\eta(t)$ is irreversible and dissipative force $F_\text{d}(t)$ is reversible. Here, we will show how such a scenario is possible.

Intuitively, one might expect that “irreversible plus reversible is irreversible”. However, once we take a closer look, it becomes clear that this is not necessarily the case. Let's call some magnitude $X(t)$ irreversible if $m,n \in \mathbb{N}$ exist so that $\Delta \langle X^m(0)X^n(t) \rangle \equiv \langle X^m(0)X^n(t) \rangle - \langle X^m(0)X^n(-t) \rangle \neq 0$. In our special case we know that, e.g., $\Delta \langle \tilde{F}(0)\tilde{F}^3(t) \rangle = 0$, $\Delta \langle F_\text{d}(0)F_\text{d}^3(t) \rangle = 0$ and $\Delta \langle \eta(0)\eta^3(t) \rangle \neq 0$. Now the general question above reduces to the specific question if this is possible, and indeed it is:
\begin{align}\label{eq:irr1}
	\Delta &\langle \tilde{F}(0)\tilde{F}^3(t) \rangle = \Delta \langle \eta(0)\eta^3(t) \rangle + 3 \Delta \langle \eta(0)\eta^2(t)F_\text{d}(t) \rangle    \nonumber \\
	&+ 3 \Delta \langle \eta(0)\eta(t)F_\text{d}^2(t) \rangle +  \Delta \langle \eta(0)F_\text{d}^3(t) \rangle + \Delta \langle F_\text{d}(0)\eta^3(t) \rangle    \nonumber \\
	&+ 3 \Delta \langle F_\text{d}(0)\eta^2(t)F_\text{d}(t) \rangle + 3 \Delta \langle F_\text{d}(0)\eta(t)F_\text{d}^2(t) \rangle   \nonumber \\
    &+ \Delta \langle F_\text{d}(0)F_\text{d}^3(t) \rangle = 0
\end{align}
If $\Delta \langle F_\text{d}(0)F_\text{d}^3(t) \rangle = 0$, Eq.~(\ref{eq:irr1}) is correct if
\begin{align}\label{eq:irr2}
	- \Delta &\langle \eta(0)\eta^3(t) \rangle = 3 \Delta \langle \eta(0)\eta^2(t)F_\text{d}(t) \rangle + 3 \Delta \langle \eta(0)\eta(t)F_\text{d}^2(t) \rangle    \nonumber \\
	&+  \Delta \langle \eta(0)F_\text{d}^3(t) \rangle + \Delta \langle F_\text{d}(0)\eta^3(t) \rangle    \\
    &+ 3 \Delta \langle F_\text{d}(0)\eta^2(t)F_\text{d}(t) \rangle + 3 \Delta \langle F_\text{d}(0)\eta(t)F_\text{d}^2(t) \rangle   \nonumber
\end{align}
It is obvious from Fig.~\ref{fig:FigSKO}(b) that Eq.~(\ref{eq:irr1}) is fulfilled there. In fact, it doesn't change much if $\Delta \langle F_\text{d}(0)F_\text{d}^3(t) \rangle$ equals zero or not. The latter adds just an additional term to Eq.~(\ref{eq:irr2}).

The case $\Delta \langle \tilde{F}(0)\tilde{F}^3(t) \rangle = 0$, $\Delta \langle F_\text{d}(0)F_\text{d}^3(t) \rangle \neq 0$ and $\Delta \langle \eta(0)\eta^3(t) \rangle = 0$, i.e., reversible noise and irreversible dissipative force, can be treated analogously. In general, “irreversible plus reversible is irreversible” is not a valid theorem.
\color{black}

\section{Methods for modification of the memory kernel}

\subsection{Iterative optimization of non-linear GLE}
\label{app:iterative}

In Sec.~\ref{sec:Iterative} we have used an iterative optimization to find memory kernels which correctly reproduce the VACF. Here, we provide some details on the methodology.

The iterative reconstruction we applied is a mixture of the original method presented in Ref.~\cite{JungHankeSchmid2017} and an improved update scheme proposed in Refs.~\cite{klippenstein2023bottom,Klippensteinetal2024}. The reconstruction is initialized with the Volterra kernel $K_0 = K_\text{V}$ as presented in Eq.~(\ref{eq:Kernel}). Subsequently, we iteratively perform the following steps:
\begin{enumerate}
    \item Perform GLE simulations using the current optimal kernel $K_i$ and Gaussian correlated noise with a correlation function proportional to the memory kernel.
    \item Extract the dynamic correlation functions from the GLE simulations and calculate the kernel $\hat{K}_i.$ Here, $\hat{K}_i$ is evaluated from Eq.~(\ref{eq:Kernel}) using the GLE correlation functions  with the only exception that instead of the oscillator force $\tilde{F}_\text{MD}$ we use the total force $F_0$.
    \item Calculate the updated kernel, $K_{i+1} = K_i \frac{1 + \hat{K}_\text{V} }{1 + \hat{K}_i}$, where $ \hat{K}_\text{V}$ is extracted using Eq.~(\ref{eq:Kernel}) based on the original MD correlation functions, again with the exception that the total force $F_0$ is used instead of  $\tilde{F}_\text{MD}$.
\end{enumerate}
Different from the original iterative method \cite{JungHankeSchmid2017}, we include a multiplicative correction and use the kernel $\hat{K}$ for the update instead of the raw correlation functions. This novel update scheme significantly improves the performance of  \cite{JungHankeSchmid2017}, in particular, it makes the usage of the ``correction'' time window obsolete. In consequence, 10 iterations are sufficient for a perfect adaptation in systems (except the ones with high barrier where convergence is not possible, see Sec.~\ref{sec:Iterative}).

The update scheme actually strongly corresponds to the Gauss-Newton method proposed in Ref.~\cite{klippenstein2023bottom}. Here, we have significantly simplified the technique, applying a constant value for regularization and using directly the memory kernel $K(t)$ instead of the integrated memory kernel $G(t) = \int_0^t K(s) \diff s$ as in Ref.~\cite{klippenstein2023bottom}. 
	
\subsection{Approximate correction of the kernel}
\label{app:ModKern}

It has been shown \cite{Klipp2021, Klippensteinetal2024, Wolf2025} that the Volterra kernel can be modified in such a way that the VACF will be correctly reproduced in GLE simulations of fully anharmonic systems. This can be done either by iterative reconstruction of the kernel (see above) or by a simple approximation formula \cite{Klipp2021}:
\begin{equation}\label{eq:Ka2}
	K_{\text{a2}}(t) = K_\text{V}(t) + \frac{\langle F_\text{e}(0)\eta(t) \rangle_\text{NS}}{\kB T}
\end{equation}
For $\langle \eta(t)F_\text{e}(0) \rangle_\text{NS}$, cf. Eqs.~\ref{eq:FDRB1} and \ref{eq:FDRB3} as well as Appendix \ref{app:BrokenFDR}.

We found empirically that inclusion of a factor $\frac{2}{3}$ gives even a better - almost perfect - approximation for our simulations:
\begin{equation}\label{eq:Ka3}
	K_{\text{a3}}(t) = K_\text{V}(t) + \frac{2}{3}	\frac{\langle F_\text{e}(0)\eta(t) \rangle_\text{NS}}{\kB T}
\end{equation}

However, both of these formulae start failing when a barrier is included in the external potential, cf. Sec. \ref{sec:Desync}.

\section{Force correlations}
\label{app:ForceCorr}

\begin{figure}[h!]
	\includegraphics[width=0.92\linewidth]{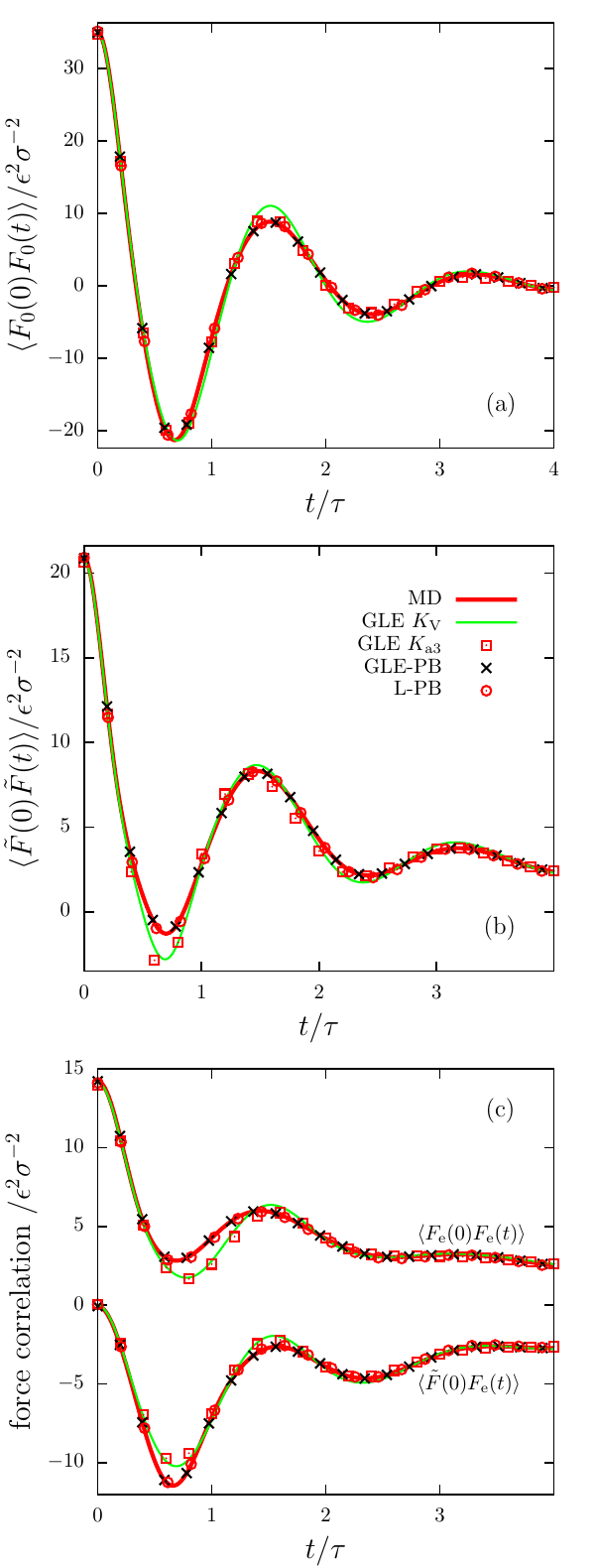}
	\caption{Force correlations for the fully anharmonic system B0A, determined via different simulation methods. (a) $\langle F_0(0)F_0(t) \rangle$. (b) $\langle \tilde{F}(0)\tilde{F}(t) \rangle$, (c) $\langle F_{\text{e}}(0)F_{\text{e}}(t) \rangle$ and $\langle \tilde{F}(0)F_{\text{e}}(t) \rangle$.}
	\label{fig:Fig_force_corr_39}
\end{figure}

In the main text, we have focused on the VACF, here we show some more results for the force correlations. Fig.~\ref{fig:Fig_force_corr_39} presents several correlations for a fully anharmonic system. Both playback methods reproduce all MD curves correctly, while GLE simulations with the Volterra kernel deviate from all of them. This is not surprising since these forces are proportional to the third power of position or distance, i.e., they are 'higher' correlations.

GLE simulations with an appropriately modified kernel $K_{\text{a3}}$ (cf. Eq.~(\ref{eq:Ka3})) reproduce only the autocorrelation of the total force $F_0$ correctly (Fig.~\ref{fig:Fig_force_corr_39}(a)), but not the others. This means that deviations in the other three correlations cancel each other out so that $\langle F_0(0)F_0(t) \rangle = \langle F_\text{e}(0)F_\text{e}(t) \rangle + 2 \langle \FGLE(0)F_\text{e}(t) \rangle + \langle \FGLE(0)\FGLE(t) \rangle$ becomes correct.

\section{Dependence of desynchronization on different parameters}
\label{app:DepInst}

\begin{figure}
	\includegraphics[width=0.85\linewidth]{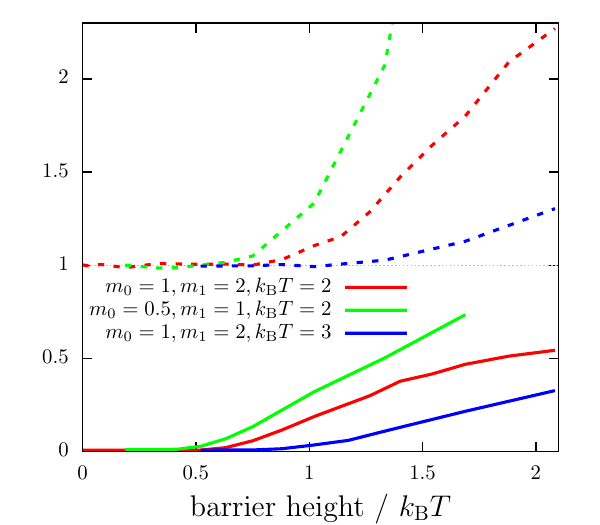}
	\caption{$\langle v_0^4 \rangle_\text{GLE-PB}/\langle v_0^4 \rangle_\text{th}$ (dashed lines) and $\sqrt{\langle \Delta ^2 v_0 \rangle/2\langle v_0^2 \rangle_\text{th}}$ (full lines) for systems like B$n$A but with different $m_0/m$, $m_1/m$ and $\kB T/\epsilon$ and various negative $a_\text{e}$, plotted vs. the barrier height. Subscript 'th' indicates the theoretical value according to the Boltzmann distribution.}
	\label{fig:Fig_Stab_mT}
\end{figure}

In the main text we have shown how the barrier height leads to a synchronization/desynchronization transition when applying GLE-PB. Here, we study empirically the dependence of desynchronization in GLE-PB on different parameters. In Sec. \ref{sec:Desync} it is shown that an increasing barrier height in the external potential clearly has a destabilizing effect. So we expect that increasing the temperature will act stabilizing. This is confirmed by blue vs. red line in Fig.~\ref{fig:Fig_Stab_mT}. There we can also see that increasing weights of particle and oscillator shift desynchronization to a somewhat higher barrier, too (red vs. green line).

\begin{figure}[h!]
	\includegraphics[width=0.85\linewidth]{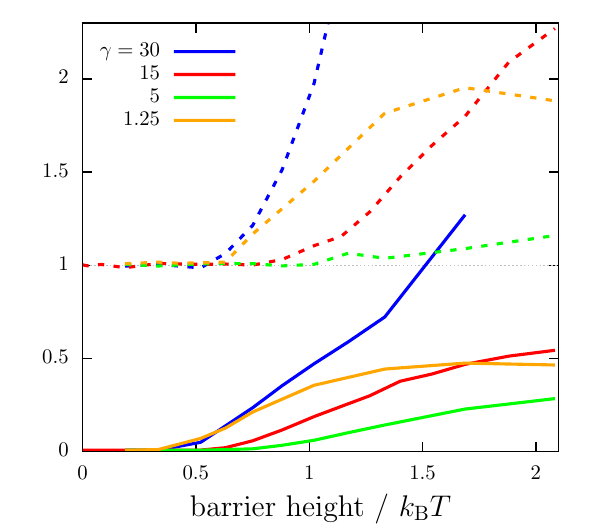}
	\caption{$\langle v_0^4 \rangle_\text{GLE-PB}/\langle v_0^4 \rangle_\text{th}$ (dashed lines) and $\sqrt{\langle \Delta ^2 v_0 \rangle/2\langle v_0^2 \rangle_\text{th}}$ (full lines) for systems like B$n$A but with different $\gamma/m \tau^{-1}$ and various negative $a_\text{e}$, plotted vs. the barrier height.}
	\label{fig:Fig_Stab_gamma}
\end{figure}

Increasing the friction coefficient $\gamma$, which acts on the oscillator, has a non-monotonic effect (cf. Fig.~\ref{fig:Fig_Stab_gamma}): decreasing it from $30 \, m\tau^{-1}$ over 15 to 5 (blue, red, green) is stabilizing. A further reduction to $\gamma = 1.25 \, m\tau^{-1}$ (orange), however, shifts desynchronization back to a lower barrier.

\begin{figure}[h!]
	\includegraphics[width=0.85\linewidth]{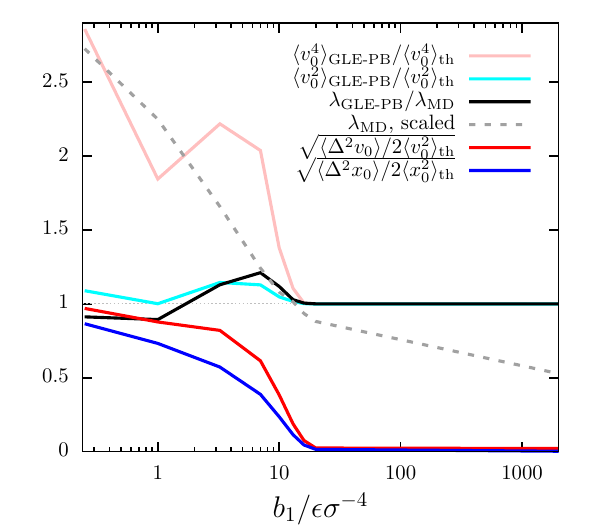}
	\caption{Different ratios and relative deviations for systems like B7A (barrier height $h = 1.02 \, \kB T$) with various $b_1$. $\lambda$ is the frequency of barrier crossing. $\langle \Delta ^2 x_0 \rangle$ and $\langle \Delta ^2 v_0 \rangle$ are the mean-square differences between GLE-PB and MD of the particle's position and velocity, respectively. The scaled $\lambda_\text{MD}$ is $\lambda_\text{MD}/2\lambda_{\text{MD},\, b_1=200000}$, i.e., half the ratio to $\lambda_\text{MD}$ for an extremely high $b_1$.}
	\label{fig:Fig_Stab_b1}
\end{figure}

Fig.~\ref{fig:Fig_Stab_b1} shows the effect of non-linear oscillator coupling for a given barrier height of $\, \approx \kB T$. If $b_1$ \small{$\gtrsim$} \normalsize{$16$}, synchronization is stable, i.e., stronger coupling is stabilizing, while the frequency of barrier crossing (gray dashed line) decreases with increasing $b_1$.

These relations are purely empirical. However, higher stability of GLE-PB with higher $T$ and $b_1$ was to be expected from Eq.~(\ref{eq:DeltaFsofort}) since $K_\text{V}(0) \propto \sqrt{b_1 \kB T}$ \cite{JungJung2023}. Nevertheless, the best practical, quick and simple method we have found for determining whether GLE-PB yields the correct results is to examine the value of $\langle v_0^4 \rangle$. If this was equal to $3 (\kB T/m_0)^2$, then all other quantities have also been correctly reproduced.

\section{Shoulders in the external potential}
\label{app:AsymPot}

So far, one might come to the conclusion that the instability of GLE-PB is caused by rare jumps of the particle over a high potential barrier. However, Eq.~(\ref{eq:DeltaFsofort}) implies that such instabilities can also occur if there is no barrier but a shoulder in the external potential. Therefore, we investigated potentials as described in Eq.~(\ref{eq:extPot}) with $a_\text{e} < 0$ for $x_0 < 0$ and $a_\text{e} = 0$ for $x_0 \geq 0$. This potential has a single minimum at $x_{0\text{m}} = -\sqrt{-a_\text{e}/b_\text{e}} < 0$ ($V_\text{e}(x_{0\text{m}}) = -\frac{1}{4} \frac{a^2_\text{e}}{b_\text{e}}$) and a shoulder at the origin.

\begin{figure}
	\includegraphics[width=0.95\linewidth]{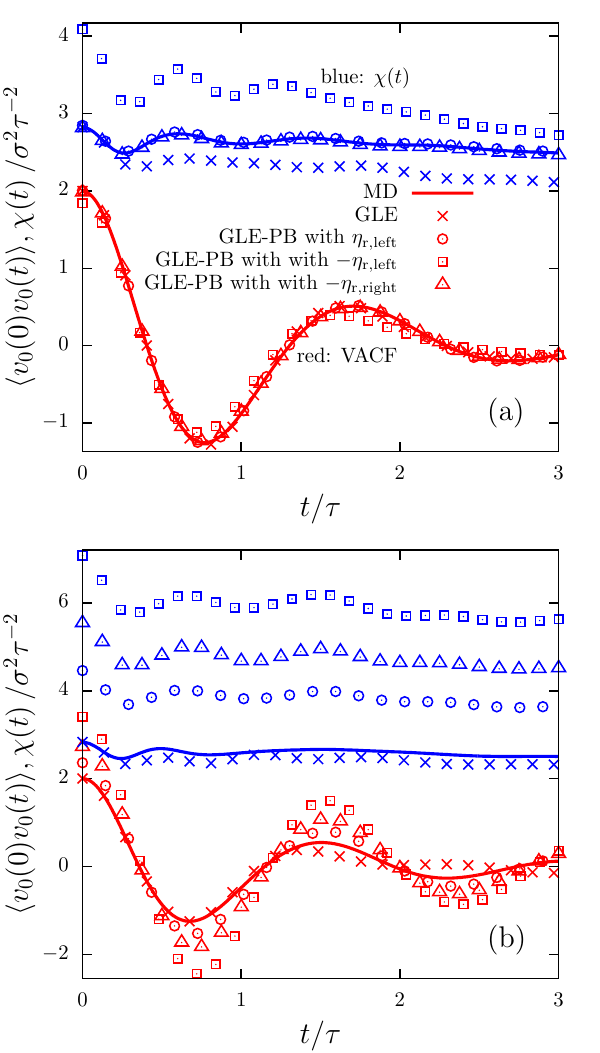}
	\caption{VACF (red) and its standard deviation $\chi (t) = \sqrt{\langle v_0^2(0)v_0^2(t) \rangle - \langle v_0(0)v_0(t) \rangle ^2}$ (blue), determined with different methods for systems with (a) low and (b) high shoulder. low: $a_\text{e}/\epsilon \sigma^{-2} = -5$ for $x_0<0$, $a_\text{e}=0$ for $x_0 \geq 0$, $\gamma/m \tau^{-1} =30$, other parameters as B$n$A. high: do., but $a_\text{e}/\epsilon \sigma^{-2} =-9$ for $x_0<0$.}
	\label{fig:Fig_schulter}
\end{figure}

 We found that for a low shoulder, GLE-PB works perfectly as can be seen in Fig.~\ref{fig:Fig_schulter}(a). For this figure, GLE-PB was conducted with different $\etaMD$: the noise recorded in the above described $V_\text{e}$ is called $\eta_{\text{MD,left}}$, and $\eta_{\text{MD,right}}$ is the noise recorded in the mirrored external potential with $x_{0\text{m}} = \sqrt{-a_\text{e}/b_\text{e}} > 0$. The VACF is described correctly if $\eta_{\text{MD,left}}$ or $-\eta_{\text{MD,right}}$ is used, but not if $-\eta_{\text{MD,left}}$ is applied. This shows that noise with inverted amplitude matches the mirrored external potential, which fits perfectly to the position dependence of noise analyzed in Sec. \ref{sec:PosDep}.
 
Even normal GLE simulations produce the VACF (almost) exactly. Its standard deviation, however, is only reproduced correctly by GLE-PB with both of the above mentioned $\etaMD$.

For a high shoulder, none of the applied methods yields the VACF and its standard deviation correctly, cf. Fig.~\ref{fig:Fig_schulter}(b). In fact, even normal GLE simulations show better results than GLE-PB.

Summarizing we can say that the effect of shoulders is basically the same as that of barriers.

	\FloatBarrier

	\bibliographystyle{unsrt}
	\bibliography{library_local.bib}

\end{document}